\documentclass[twocolumn]{aastex631}

\usepackage{txfonts}

\newcommand{\cm}{~cm$^{-1}$}
\newcommand{\um}{~$\mu$m}

\newcommand{\exocross}{\textsc{ExoCross}}

 \shorttitle{Individual vibrational band spectroscopy for WASP-33b}
\shortauthors{Wright et al.}
\graphicspath{{./}{figures/}}

\begin{document}

\title{A spectroscopic thermometer: individual vibrational band spectroscopy with the example of OH in the atmosphere of WASP-33b}

\author[0000-0002-3852-9562]{Sam O.M. Wright}
\affiliation{Department of Physics and Astronomy, University College London, Gower Street, WC1E 6BT London, United Kingdom}
\email{swright@star.ucl.ac.uk}

\author[0000-0003-4698-6285]{Stevanus K. Nugroho}
\affiliation{Astrobiology Center, NINS, 2-21-1 Osawa, Mitaka, Tokyo 181-8588, Japan}
\affiliation{National Astronomical Observatory of Japan, NINS, 2-21-1 Osawa, Mitaka, Tokyo 181-8588, Japan}

\author[0000-0002-7704-0153]{Matteo Brogi}
\affiliation{Dipartimento di Fisica, Universit\`a degli Studi di Torino, via Pietro Giuria 1, I-10125, Torino, Italy}
\affiliation{Department of Physics, University of Warwick, Coventry CV4 7AL, UK}
\affiliation{INAF-Osservatorio Astrofisico di Torino, Via Osservatorio 20, I-10025 Pino Torinese, Italy}

\author[0000-0002-9308-2353]{Neale P. Gibson}
\affiliation{School of Physics, Trinity College Dublin, The University of Dublin, Dublin 2, Ireland}

\author[0000-0001-6391-9266]{Ernst J. W. de Mooij}
\affiliation{Astrophysics Research Centre, Queen's University Belfast, Belfast BT7 1NN, UK}

\author[0000-0002-4205-5267]{Ingo Waldmann}
\affiliation{Department of Physics and Astronomy, University College London, Gower Street, WC1E 6BT London, United Kingdom}

\author[0000-0002-4994-5238]{Jonathan Tennyson}
\affiliation{Department of Physics and Astronomy, University College London, Gower Street, WC1E 6BT London, United Kingdom}

\author[0000-0003-3309-9134]{Hajime Kawahara}
\affiliation{Department of  of Space Astronomy and Astrophysics, Institute of Space and Astronautical Science, Japan Aerospace Exploration Agency, 3-1-1, Yoshinodai, Chuo-ku, Sagamihara, Kanagawa, 252-5210, Japan}

\author[0000-0002-4677-9182]{Masayuki Kuzuhara}
\affiliation{Astrobiology Center, NINS, 2-21-1 Osawa, Mitaka, Tokyo 181-8588, Japan}
\affiliation{National Astronomical Observatory of Japan, NINS, 2-21-1 Osawa, Mitaka, Tokyo 181-8588, Japan}

\author[0000-0003-3618-7535]{Teruyuki Hirano}
\affiliation{Astrobiology Center, NINS, 2-21-1 Osawa, Mitaka, Tokyo 181-8588, Japan}
\affiliation{National Astronomical Observatory of Japan, NINS, 2-21-1 Osawa, Mitaka, Tokyo 181-8588, Japan}
\affiliation{Department of Astronomy, School of Science, The Graduate University for Advanced Studies (SOKENDAI), 2-21-1 Osawa, Mitaka, Tokyo 181-8588, Japan}

\author[0000-0001-6181-3142]{Takayuki Kotani}
\affiliation{Astrobiology Center, NINS, 2-21-1 Osawa, Mitaka, Tokyo 181-8588, Japan}
\affiliation{National Astronomical Observatory of Japan, NINS, 2-21-1 Osawa, Mitaka, Tokyo 181-8588, Japan}
\affiliation{Department of Astronomy, School of Science, The Graduate University for Advanced Studies (SOKENDAI), 2-21-1 Osawa, Mitaka, Tokyo 181-8588, Japan}

\author[0000-0003-3800-7518]{Yui Kawashima}
\affiliation{Cluster for Pioneering Research, RIKEN, 2-1 Hirosawa, Wako, Saitama 351-0198, Japan}

\author[0000-0003-1298-9699]{Kento Masuda}
\affiliation{Department of Earth and Space Science, Osaka University, Osaka 560-0043, Japan}

\author[0000-0002-4125-0140]{Jayne L. Birkby}
\affiliation{Astrophysics, Department of Physics, University of Oxford, Keble Road, Oxford OX1 3RH, UK}

\author[0000-0002-9718-3266]{Chris A. Watson}
\affiliation{School of Mathematics and Physics, Queen's University Belfast, University Road, Belfast, BT7 1NN, United Kingdom}

\author[0000-0002-6510-0681]{Motohide Tamura}
\affiliation{Department of Astronomy, Graduate School of Science, The University of Tokyo, 7-3-1 Hongo, Bunkyo-ku, Tokyo 113-0033, Japan}
\affiliation{Astrobiology Center, NINS, 2-21-1 Osawa, Mitaka, Tokyo 181-8588, Japan}
\affiliation{National Astronomical Observatory of Japan, NINS, 2-21-1 Osawa, Mitaka, Tokyo 181-8588, Japan}

\author[0000-0001-9229-8315]{Konstanze Zwintz}
\affiliation{Institute for Astro- and Particle Physics, University of Innsbruck, Technikerstrasse 25/8, A-6020 Innsbruck, Austria}

\author[0000-0002-7972-0216]{Hiroki Harakawa}
\affiliation{Subaru Telescope, 650 N. Aohoku Place, Hilo, HI 96720, USA}

\author[0000-0002-9294-1793]{Tomoyuki Kudo}
\affiliation{Subaru Telescope, 650 N. Aohoku Place, Hilo, HI 96720, USA}

\author{Klaus Hodapp}
\affiliation{University of Hawaii, Institute for Astronomy, 640 N. Aohoku Place, Hilo, HI 96720, USA}

\author{Shane Jacobson}
\affiliation{University of Hawaii, Institute for Astronomy, 640 N. Aohoku Place, Hilo, HI 96720, USA}

\author{Mihoko Konishi}
\affiliation{Faculty of Science and Technology, Oita University, 700 Dannoharu, Oita 870-1192, Japan}

\author{Takashi Kurokawa}
\affiliation{Astrobiology Center, NINS, 2-21-1 Osawa, Mitaka, Tokyo 181-8588, Japan}
\affiliation{Institute of Engineering, Tokyo University of Agriculture and Technology, 2-24-16, Nakacho, Koganei, Tokyo, 184-8588, Japan}

\author[0000-0001-9326-8134]{Jun Nishikawa}
\affiliation{National Astronomical Observatory of Japan, NINS, 2-21-1 Osawa, Mitaka, Tokyo 181-8588, Japan}
\affiliation{Astrobiology Center, NINS, 2-21-1 Osawa, Mitaka, Tokyo 181-8588, Japan}
\affiliation{Department of Astronomy, School of Science, The Graduate University for Advanced Studies (SOKENDAI), 2-21-1 Osawa, Mitaka, Tokyo 181-8588, Japan}

\author[0000-0002-5051-6027]{Masashi Omiya}
\affiliation{Astrobiology Center, NINS, 2-21-1 Osawa, Mitaka, Tokyo 181-8588, Japan}
\affiliation{National Astronomical Observatory of Japan, NINS, 2-21-1 Osawa, Mitaka, Tokyo 181-8588, Japan}

\author{Takuma Serizawa}
\affiliation{Institute of Engineering, Tokyo University of Agriculture and Technology, 2-24-16, Nakacho, Koganei, Tokyo, 184-8588, Japan}
\affiliation{National Astronomical Observatory of Japan, NINS, 2-21-1 Osawa, Mitaka, Tokyo 181-8588, Japan}

\author{Akitoshi Ueda}
\affiliation{Astrobiology Center, NINS, 2-21-1 Osawa, Mitaka, Tokyo 181-8588, Japan}
\affiliation{National Astronomical Observatory of Japan, NINS, 2-21-1 Osawa, Mitaka, Tokyo 181-8588, Japan}
\affiliation{Department of Astronomy, School of Science, The Graduate University for Advanced Studies (SOKENDAI), 2-21-1 Osawa, Mitaka, Tokyo 181-8588, Japan}

\author[0000-0003-4018-2569]{Sébastien Vievard}
\affiliation{Astrobiology Center, NINS, 2-21-1 Osawa, Mitaka, Tokyo 181-8588, Japan}
\affiliation{Subaru Telescope, 650 N. Aohoku Place, Hilo, HI 96720, USA}

\author[0000-0001-9286-9501]{Sergei N. Yurchenko}
\affiliation{Department of Physics and Astronomy, University College London, Gower Street, WC1E 6BT London, United Kingdom}



\begin{abstract}
Individual vibrational band spectroscopy presents an opportunity to examine exoplanet atmospheres in detail by distinguishing where the vibrational state populations of molecules differ from the current assumption of a Boltzmann distribution. Here, retrieving vibrational bands of OH in exoplanet atmospheres is explored using the hot Jupiter WASP-33b as an example. We simulate low-resolution spectroscopic data for observations with the JWST's NIRSpec instrument and use high resolution observational data obtained from the Subaru InfraRed Doppler instrument (IRD). Vibrational band-specific OH cross section sets  are constructed and used in retrievals on the (simulated) low and (real) high resolution data. Low resolution observations are simulated for two  WASP-33b emission scenarios: under the assumption of local thermal equilibrium (LTE) and a toy non-LTE model for vibrational excitation of selected bands.
We show that mixing ratios for individual bands can be retrieved with sufficient precision to allow the vibrational population distributions of the forward models to be reconstructed. A simple fit for the Boltzmann distribution in the LTE case shows that the vibrational temperature is recoverable in this manner. For high resolution, cross-correlation applications, we apply the individual vibrational band analysis to an IRD  spectrum of WASP-33b, applying an `un-peeling' technique. Individual detection significances for the two strongest bands are shown to be in line with Boltzmann distributed vibrational state populations consistent with the effective temperature of the WASP-33b atmosphere reported previously. We show the viability of this approach for analysing the individual vibrational state populations behind observed and simulated spectra including reconstructing state population distributions. 

\end{abstract}



\section{Introduction} \label{sec:intro}
The study of exoplanet atmospheres is motivated by a desire to understand the physical properties of exoplanets and their origins. These properties reflect a range of planetary environments which can differ drastically from those in our Solar System and are yet to be fully understood; to gain a better understanding we must determine their atmospheric composition and thermal structure. We can also extract valuable information about the origins of these planets and their formation histories, including information about planetary migration and providing constraints on the modelling of protoplanetary disks \citep{11ObMuBe}. Beyond enhancing our current understanding, characterising exoplanets can inform the direction of future research, targets for follow-up study and the priorities for designing future space missions and ground-based facilities.

We deduce these characteristics via the process of retrieval; it allows us to infer parameters related to the atmosphere's composition and structure from spectroscopic data acquired by observing the planet. Spectra are produced based on atmospheric models of physical processes taking place and the radiative transfer of photons through the planet's atmosphere. Statistical techniques (typically Bayesian inference methods) are used to match the modelled spectra with the observed spectral data given models produced with the chosen parameterisation. Optimal values for the parameters are found via these methods, where optimal is quantified as the maximum a posteriori probability.

Typically the temperature profiles in exoplanetary atmospheric retrievals are obtained via the fitting of a complex atmospheric model to an observed absorption (transit) or emission spectrum, where the temperature is one of the factors shaping the spectral profile. Most notably, temperature affects the magnitude (intensity) of different absorption/emission bands as well as their shapes \citep[e.g.][]{13TeTiSa.exo}. 
Here we explore an alternative, `spectroscopic thermometer' approach based on the relative intensities of individual molecular bands as an estimate  of their vibrational populations and therefore the corresponding vibrational temperature. 
Knowledge of vibrational state populations obtained via band-by-band retrievals provides an opportunity to infer the molecular vibrational temperature directly from their spectral features. Additionally, any outlying band population would indicate the molecule's departure from the local thermal equilibrium (LTE)  within the planet's atmosphere. Especially favourable for this purpose are the (hot) bands of some molecules which are characterised by  substantial  displacement from each other within a relatively small spectroscopic region \citep{22WrWaYu.nLTE}; typical examples are diatomics and linear polyatomics. The ability to analyse molecular spectra with such granularity will play an increasingly important role in the characterization of exoplanet atmospheres as accounting for divergences from local thermodynamic equilibrium and could greatly impact our understanding of atmospheric conditions, such as temperature structure \citep{21FoYoSh.nLTE,19FiHexx.nLTE} and wind characteristics \citep{21BoFoKo.nLTE}.

While the approach outlined here naturally enables non-LTE analyses, it also provides advantages for LTE atmospheric analysis with high and low resolution observations. For high resolution data it allows temperature measurement without needing to run computationally expensive T-P profile retrievals. When analysing low resolution data, this approach can serve as a separate temperature diagnostic tool with the potential to avoid degeneracies in temperature retrieval results.

Here we consider the hydroxyl radical (OH) in the atmosphere of WASP-33b both for low resolution and high resolution applications. 
OH has recently been detected in the atmosphere of WASP-33b using the cross correlation technique \citep{21NuKaHa.OH} with data from the Subaru telescope's InfraRed Doppler instrument \citep[IRD; R\,$\approx$\,70,000, ][]{Tamura2012, Kotani2018} in the spectral region of 0.97--1.75~\um\ which was later confirmed by \citet{Cont2022} using CARMENES data. This near infrared region contains the OH emission bands formed by the $\Delta v=2$ rotation-vibrational transitions from vibrational states populated at the relevant temperature of the atmosphere. The OH spectrum  has good wavelength separation between hot vibrational bands. In this work we aim to use this feature to disentangle  individual band intensities  in the atmospheric retrievals in order to target their individual vibrational populations, i.e. to  use information beyond the aggregated band level information found in the molecular data typically employed for exoplanet atmospheric analyses. We will  show that the  separation between the OH bands in the NIR is sufficient for the bands to be individually retrievable even with the current observation technology, despite the usual degeneracy of the atmospheric factors in a typical retrieval of the atmospheres of hot Jupiters. While spectral signatures have been shown to differ in non-LTE for other molecules \citep{22WrWaYu.nLTE}, it is also worth investigating OH in particular when considering non-LTE in exoplanet atmospheres since it has a well documented airglow effect in Earth's atmosphere \citep{20NoWiGo.OH, 50Mexxxx.OH} and the Martian atmosphere \citep{13ClSaGa.OH}. In non-LTE circumstances, vibrational state populations diverge substantially from a Boltzmann distribution \citep[as shown in ][]{21ChHuGu.OH} which makes the selection of OH particularly relevant for this analysis.

\section{NIR spectrum of OH  and cross section generation} 
\label{sec:xsecs}

Our goal is to investigate the possibility of  retrieving individual vibrational populations of OH in the NIR emission spectrum of a hot-Jupiter like  atmosphere, using WASP-33b as an example. The MoLLIST line list \citep{16BrBeWe.OH} for OH is used, along with the \exocross\ software \citep{Exocross} for molecular cross section generation. The software provides two capabilities crucial to this investigation; the means to filter individual vibrational bands (i.e select isolated bands while still assuming a Boltzmann distribution under the LTE condition) and also to supply custom state populations from which cross sections can be calculated. 

\begin{figure}[!ht]
    \centering
    \includegraphics[width=\columnwidth]{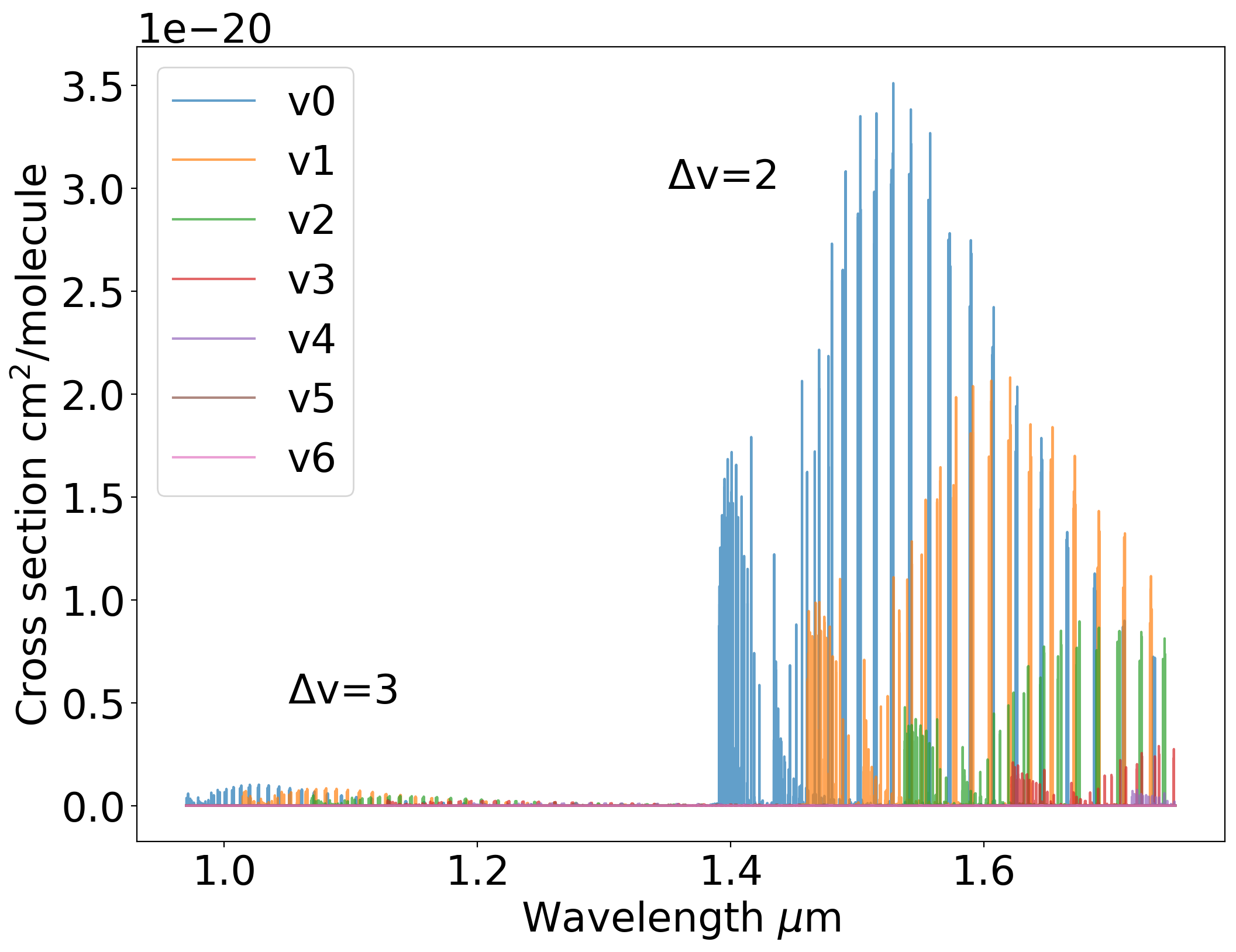}
    \caption{Individual band contributions for cross sections of OH in LTE with the NIR ($\Delta v=2$ and $\Delta v = 3$) shown for the 3050~K, 0.01~bar point on the temperature pressure grid.}
    \label{fig:filtered_xsecs}
\end{figure}

The  NIR spectrum of OH is mainly composed of the $\Delta v= 2$ vibrational bands, including the overtone  band $(2,0)$, as well as the  weaker  $\Delta v = 3$ bands, see Fig.~\ref{fig:filtered_xsecs} where an LTE spectrum of OH for the equilibrium temperature of $T=3050$~K  is shown, with the individual bands ($\Delta v = 2$ and $\Delta v = 3$) depicted in different colours. Here, assuming the LTE conditions, cross sections of 7 (in the first instance) individual vibrational bands ($0 \le v''\le 6$) of OH were generated using a grid spacing of 0.03\cm; this grid spacing corresponds to a minimum resolving power of $R\approx190,000$ (at 1.75~\um) which allowed us to retain flexibility at the modelling stage and the potential for applications of the cross section set to higher resolution data in the future. The constant grid spacing allows us to guarantee that we did not introduce artefacts from the grid subdivision performed by \exocross\ to achieve a fixed resolution across the wavelength range. The vibrational bands were individually isolated via the filtration method where \exocross\ selects only those spectroscopic lines originating from the desired vibrational state.  In line with the LTE approximation, their magnitudes  drop  due to the Boltzmann factor as well as the corresponding vibrational band strength. For reference,  we also  retained the standard set of cumulative  cross sections of OH  for modelling the contribution of OH in LTE to the spectrum of WASP-33b.
Although our goal is to examine the effects in emission, we remain consistent with atmospheric codes by generating cross sections for the absorption case. Atmospheric modelling codes (such as TauREx 3 \citep{TauRex3}) ingest absorption data and model the emission case using these absorption data equivalently by Kirchoff's law under the assumption of LTE.

In order to be able to retrieve populations of individual bands of OH in atmospheric spectroscopic models, individual band cross sections were generated with normalised populations  (state population set to 1) as  a sum of all line intensities for transitions $f\gets i$ from a given vibrational band $(v',v'')$ \citep{Exocross} 
\begin{equation}
\label{e:sigma}
\sigma(\tilde{\nu})_{v',v''} = \sum_{i,f} f_{if}(\tilde{\nu}) \frac{A_{if}^{(v',v'')} N_{\rm r-v}  }{8\pi c^2\tilde{\nu}_{if}^2} \left(1-e^{-hc\tilde{\nu}_{if}/kT}\right),
\end{equation}
where $f_{if}(\tilde{\nu})$ is a line profile, $A_{if}^{(v',v'')}$ is the corresponding Einstein A coefficient and $N_{\rm r-v}$ is a population of the lower  rotation-vibration state $(v",i)$.

In order to separate the  vibrational and rotational populations  in our analysis,  
we use the following approximation  of a product of these contributions: 
\begin{equation}
N_{\rm r-v} = N_v N_{\rm rot}
\end{equation}
where  $N_v $ is a population of the vibrational (or vibronic) state $v$ and $N_{\rm rot}$ is the population of the rotational state. For the rotational population a  Boltzmann distribution is assumed with the rotational temperature $T_{\rm rot}$ the same as structural temperature of the atmosphere as defined by the atmospheric T-P profile, taken for calculating the scale height. This approximation is similar to the Treanor distribution where the rotational and vibrational components are separated as discussed in \citet{68TrRiRe.nLTE}. The Treanor distribution takes the product of two Boltmann distributions; one in terms of the rotational temperature and one in terms of the vibrational temperature. We make this assumption for the temperature in Eq.~(\ref{e:sigma}) such that $T = T_{\rm rot}$. The effect of the vibrational population   is to scale each of the band intensities by the population of the state $v$. For example, the vibrational Boltzmann population is obtained using the formula,
\begin{equation}
\label{e:Nv}
N_{v} = \frac{e^{-E_v/k T_{\rm vib}}}{Q_v(T)}
\end{equation}
where $E_v$ is the vibrational energy of the state, $Q_v(T)$ is the temperature-dependent vibrational partition function and $k$ is the Boltzmann constant. We can relate this back to Fig.~\ref{fig:filtered_xsecs} which shows this for the circumstance where $T_{\rm vib} = T_{\rm rot}$, i.e. under the LTE condition. In this work we aim to retrieve the populations $N_{v}$ directly from the exoplanetary spectrum as an independent measure of the atmospheric temperature via $T_{\rm vib}$.

\begin{figure}[!ht]
    \centering
    \includegraphics[width=\columnwidth]{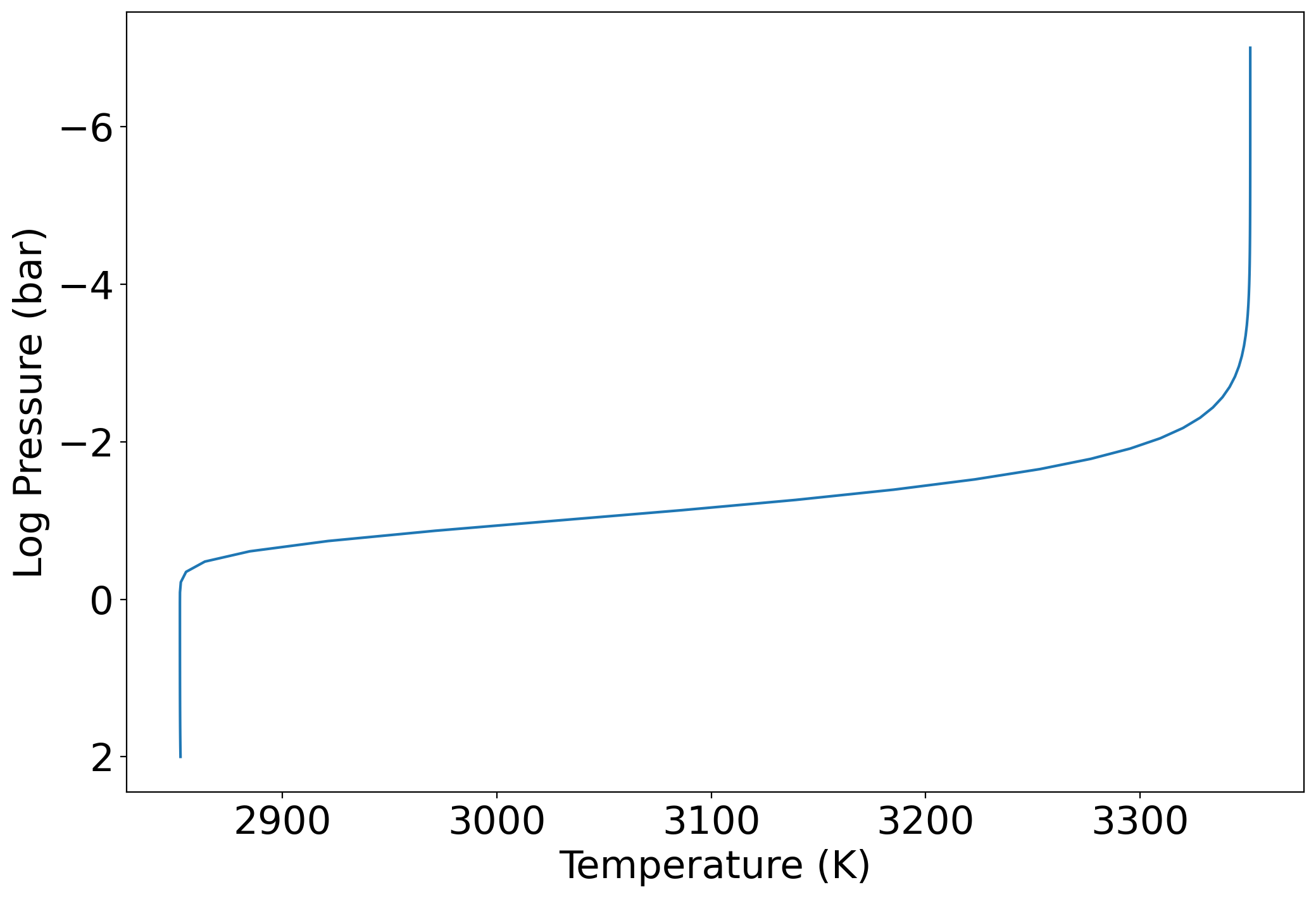}
    \caption{The Guillot T-P profile adopted for modelling the atmosphere of WASP-33b.}
    \label{fig:WASP_33b_TP_Profile}
\end{figure}

\begin{figure}
    \centering
    \includegraphics[width=0.848\columnwidth]{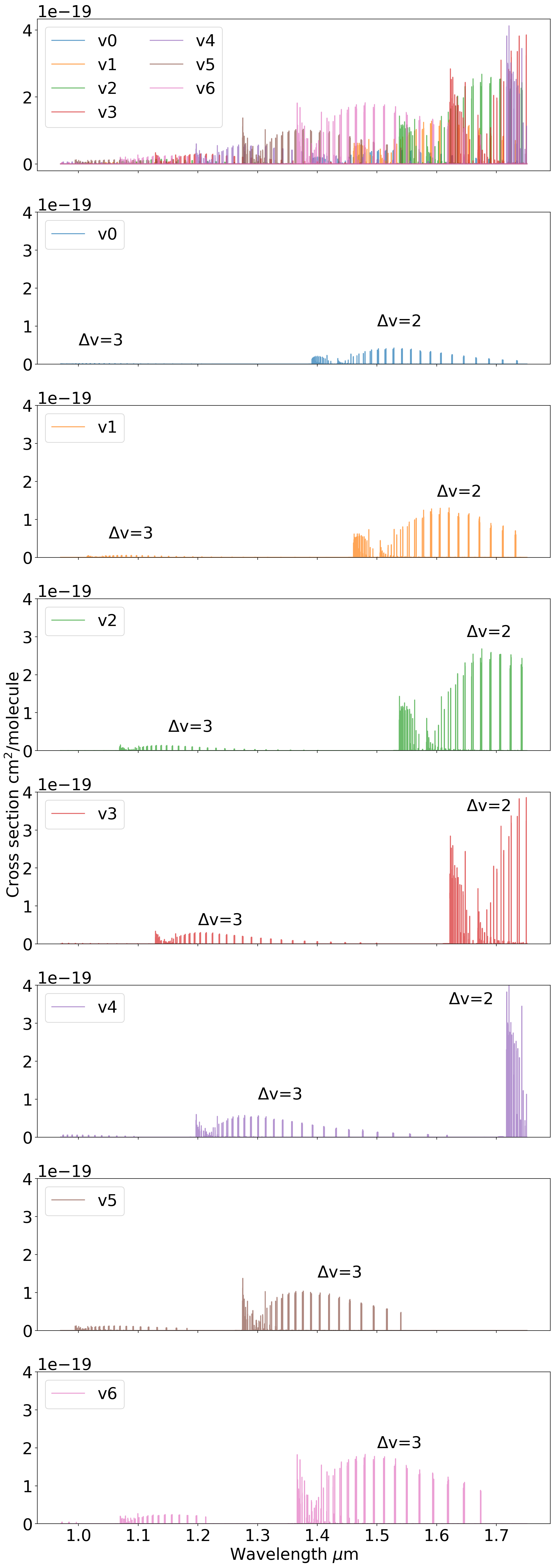}
    \caption{NIR cross sections of OH ($\Delta v=2$ and $\Delta v=3$) produced from the completely populated states plotted overlaid in the top pane for scale and individually below, all at a temperature of 3050~K and a pressure of 0.01~bar. (Note that as we progress to the final two cross section sets the $\Delta v=4$ and $\Delta v=5$ bands start to become visible)}

    \label{fig:pop1d_xsecs}
\end{figure}

At this step, we use \exocross\ to generate cross sections for OH  with bespoke populations $N_{\rm r-v}$  for each of the states defined in the MoLLIST OH line list. The \exocross\ software applies these populations for  each state with the corresponding vibrational quantum number. Individual vibrational populations can be provided in the \exocross\ program in order to isolate each desired band; in each case the vibrational population $N_{v}$ was set to 1 for the desired state and 0 for the remaining states and  a partition function is inferred. Using this method we produced seven sets of cross sections for each of the lower states ($v''=0, 1, 2,3, 4, 5, 6$) of the hydroxyl radical for the $\Delta v=2$ and $\Delta v=3$ bands; i.e. those where opacity falls into the NIR window $<1.8$~$\mu$m.  The corresponding  population of each band was normalised to 1 for temperatures ranging between 2850~K and 3350~K and pressures between $1\times 10^{-8}$~bar and $100$~bar which covers the T-P profile of WASP-33b which follows a \citet{10Guxxxx.exo} profile as shown in Fig.~\ref{fig:WASP_33b_TP_Profile}.  
The resulting cross sections are plotted in Fig.~\ref{fig:pop1d_xsecs} for a rotational temperature of $T_{\rm rot}=3050$~K and $N_v = 1.0$, where individual spectral components of the hydroxyl radical are shown; all the bands are overlaid in the top pane to show their relative intensities and then plotted individually below for clarity.

Bands with different lower states excitations $v'' \le 5$ are 
indicated by different colours. The stronger features on the right are from $\Delta v = 2$ bands, while the  weaker features on the left are  from the $\Delta v = 3$ bands.  
The labels $v0$, $v1$, $v2$, \ldots  indicate the lower vibrational states 0, 1, 2, \ldots, respectively.

For this case,  we see that the strengths of the $\Delta v = 2$ bands increase with $v$ quadratically, using an extension of the double harmonic approximation to the overtone bands \citep{21ClYu.add}, while the  intensity increase of the $\Delta v=3$ band is approximately cubic \citep{22PaClYu}. 
Due to the relative line strengths, the $\Delta v = 3$ sequence  will be  far less significant than the $\Delta v = 2$ band features until the $v5$ ($v'' = 5$) and $v6$ ($v'' = 6$) systems. The weaker $\Delta v = 3$ bands are also included as  present within our chosen spectral range. As the $\Delta v= 2$ moves away from the window, for  $v''>4$, only $\Delta v= 3$ bands remain visible. 

To come a full circle for the sake of illustration, the original LTE spectrum at a given $T$ can be then reconstructed from these $N_{v}=1$ vibrational band-isolated cross sections by scaling the individual band cross sections for given $ T = T_{\rm rot}$ to the actual population number $N_v$ in Eq.~(\ref{e:Nv}). For example,  the cross sections in Fig.~\ref{fig:filtered_xsecs} were obtained by scaling the cross sections from Fig.~\ref{fig:pop1d_xsecs} using $N_v$ from Eq.~(\ref{e:Nv}). Accordingly, there is an additional drop in their intensities with increasing $v$.

\section{Forward Modelling} \label{sec:fwd_modelling}

Forward modelling of the emission spectrum of WASP-33b  was performed by taking the requisite input parameters from \citet{21NuKaHa.OH},  listed in Table~\ref{WASP_33b_Params_Table}. These planetary and stellar parameters were used along with the generated cross sections and the atmospheric modelling code TauREx 3 \citep{TauRex3} to produce atmospheric models. The standard, unfiltered cross sections were used to generate a forward model of OH in WASP-33b's atmosphere as shown in Fig.~\ref{fig:obs_sims}; this was performed with the raw OH cross sections, allowing the retention of the native high resolution models for high resolution cross correlation analysis and the later downbinning during the observation simulations for the low resolution space borne analysis.

\begin{table}
	\footnotesize
	\centering
		\begin{tabular} {c | c }
			\hline
			\hline
			Parameter & WASP-33b Value\\
			\hline
			Planet Radius ($R_{J}$) & 1.679   \\
			Planet Mass ($M_{J}$) & 3.266 \\
			Stellar Temperature (K) & 7400\\
			Stellar Radius ($R_{\odot}$) & 1.509\\
			Stellar Metallicity (dex) & 0.1\\
			Semi-major Axis (AU) & 0.02558\\
			Period (Days) & 1.21987\\  
			\hline
			OH Volume Mixing Ratio (VMR) Profile Parameters \\
			\hline
			Surface VMR & 7.94328x10$^{-7}$\\  
			Middle VMR & 6.30957x10$^{-5}$\\
			Top VMR & 1x10$^{-10}$\\
		    Surface Pressure (Pa) & 1x10$^{7}$\\  
			Middle Pressure (Pa) & 3x10$^{3}$\\
			Top Pressure (Pa) & 1x10$^{-3}$\\
			\hline
			\hline
	\end{tabular}
	\caption{Planetary and stellar parameters used in forward modelling WASP-33b's atmosphere. 	\label{WASP_33b_Params_Table}
}
\end{table}  

\section{James Webb observation simulations and retrievals}
\subsection{Observation simulations}

\begin{figure}[!ht]
    \centering
    \includegraphics[width=\columnwidth]{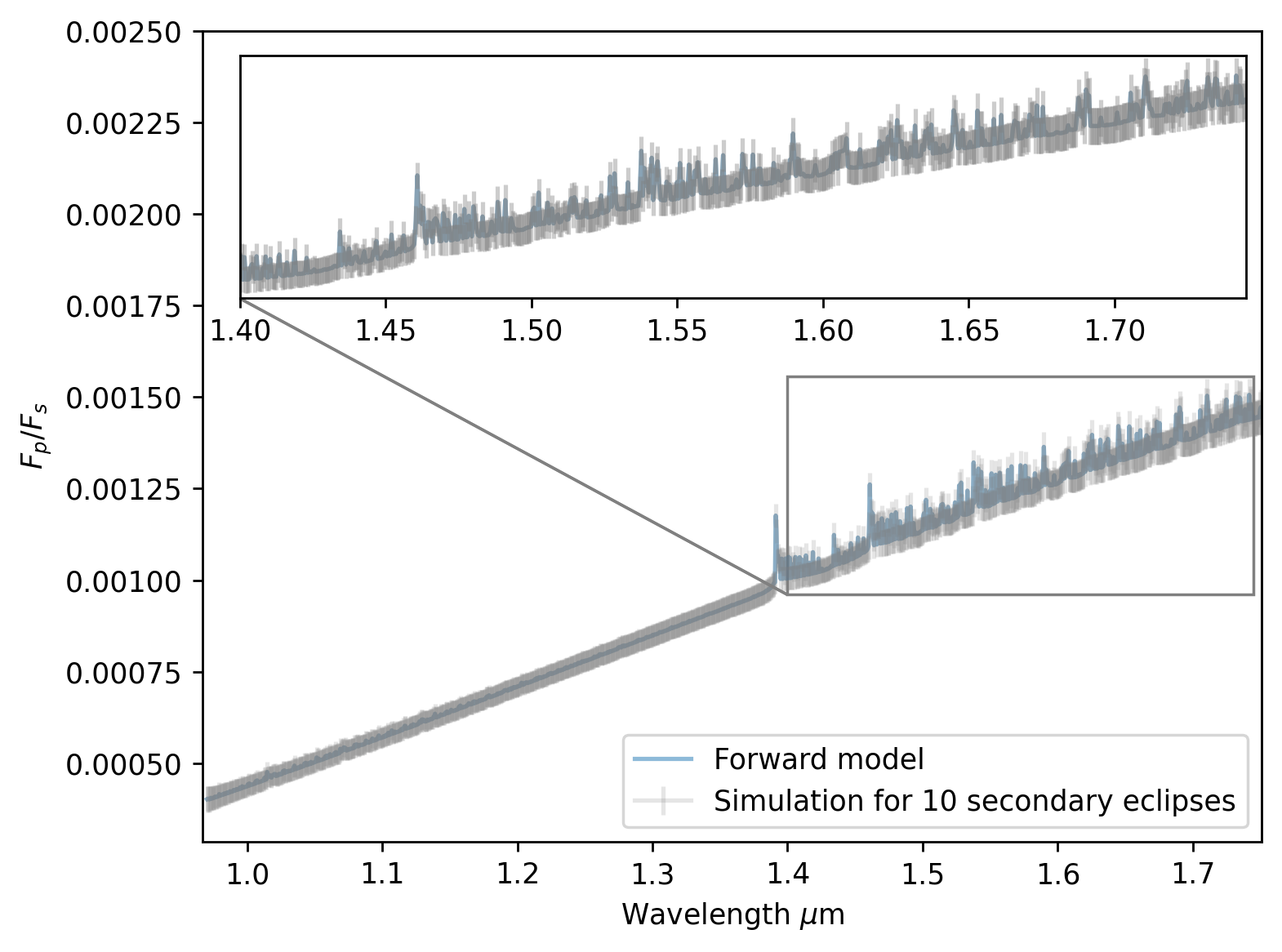}
    \caption{Observation simulation points and errors of the  WASP-33b emission spectra plotted over the forward model for observations over 10 secondary eclipses.}
    \label{fig:obs_sims}
\end{figure}

Simulations of observations of WASP-33b's atmosphere with JWST using the G140H configuration of the NIRSpec instrument were conducted  using ExoWebb \citep{Terminus} noise simulation along with Pandeia data \citep{PANDEIA} incorporated into TauREx's instrument simulation code. The simulation is shown  in Fig.~\ref{fig:obs_sims} for a 10 secondary eclipse simulation of WASP-33b, overlaying the TauREx forward model for comparison. An increasing number of simulated secondary eclipses yielded simulated observations with successively decreasing noise in the relative flux values. The 10 secondary eclipse simulation shows the visibility of OH lines clearly in the wavelength region above 1.4~\um.  
\subsection{Retrievals of vibrational populations}

Atmospheric retrievals of the populations of individual bands were performed with TauREx using a new non-LTE plugin (Wright (in prep.)) on the simulated observation data using cross sections constructed as detailed in Section \ref{sec:xsecs}. Nested sampling was performed to fit for the log of the mixing ratios for the cross sections included in each case (log$({\rm OH}n)$ with uniform priors between -12 and -1) and also the planetary radius ($R_{p}$ with uniform priors between 1.5 and 2 $R_{J}$).The fit for planetary radius is included to ensure any degeneracies in fitting are visible. From this a derived mean molecular weight ($\mu$) is calculated and its distribution displayed alongside the inferred probability distributions; this is included as a summary characteristic of the retrieved atmosphere.

The first set of retrievals was conducted using the cross sections corresponding to the vibrational Boltzmann LTE case as our baseline, where each band is effectively considered as an individual molecule with the population pre-defined using the Boltzmann LTE rule in Eq.~(\ref{e:Nv}) with $T_{\rm vib} = T_{\rm rot}$ following the T-P profile discussed in Section~\ref{sec:xsecs}. 
This can be directly correlated with a population of an upper vibrational state $v'$ or equivalently of a  lower state  $v''$, since we only deal with states differing by $\Delta v = 2$ or $\Delta v = 3$.  

Based on this property, in the following,  the vibrational bands in question $v0$, $v1$, $v2$ etc. as labelled in Fig.\ref{fig:pop1d_xsecs}  will be referenced as bands 0, 1, 2 etc., i.e. using the corresponding lower state $v''$ vibrational quantum number. 

The mixing ratios for each band associated with $v''$  were considered  in two ways: (i) following the three point mixing profile used in the forward modelling step and (ii) also using a constant mixing ratio. Though given the altitude of the thermal gradient modelled, the lower and upper points of the mixing profile cannot be well constrained on account of the limited pressure region probed by this wavelength, as such we opted to proceed to perform retrievals with a constant mixing ratio profile. 

\begin{figure}[!ht]
    \centering
    \includegraphics[width=\columnwidth]{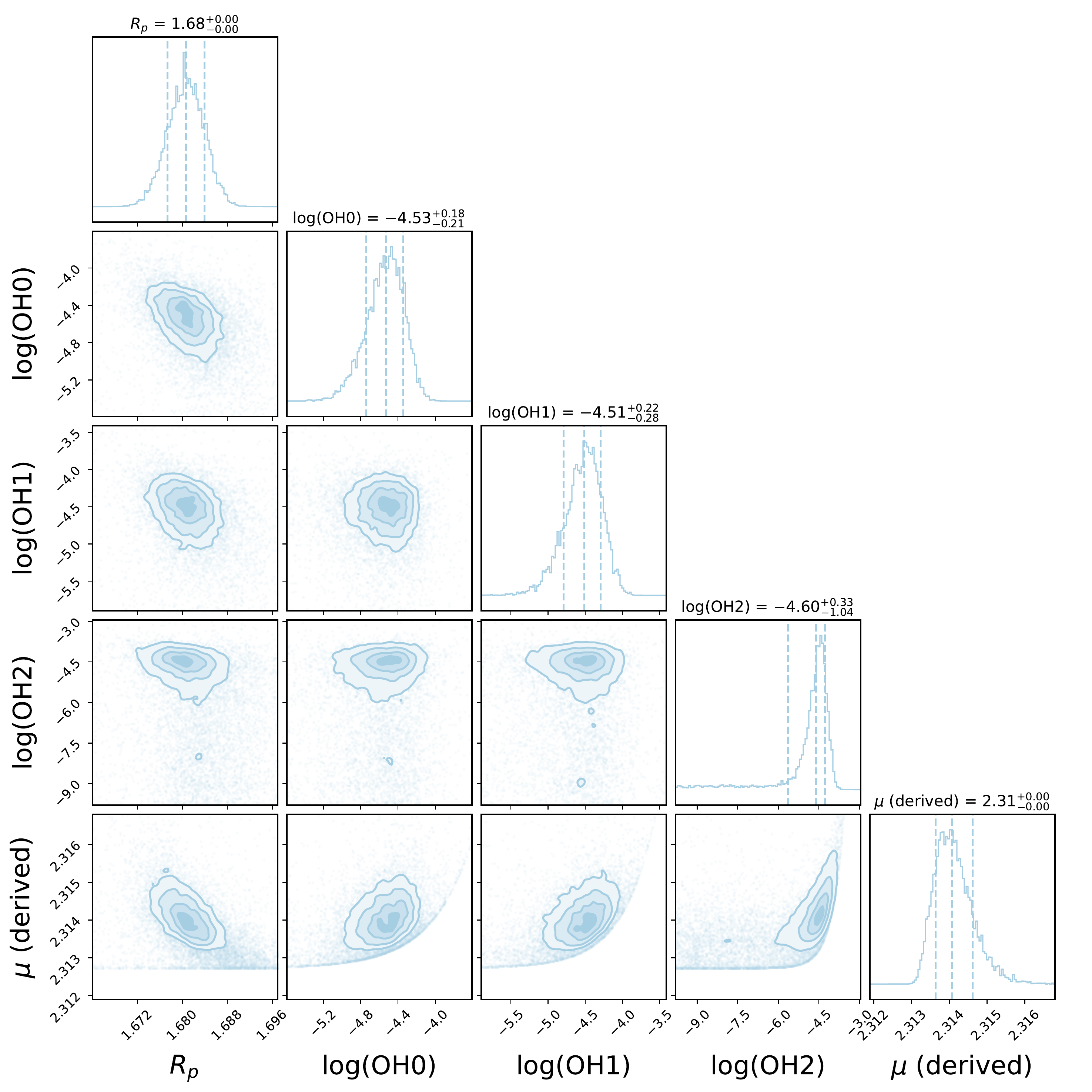}
    \caption{Retrieval nested sampling posteriors for one simulated secondary eclipse with bands 0 to 2.}
    \label{fig:0_2_1_constant_mix_posteriors}
\end{figure}

\begin{figure}[!ht]
    \centering
    \includegraphics[width=\columnwidth]{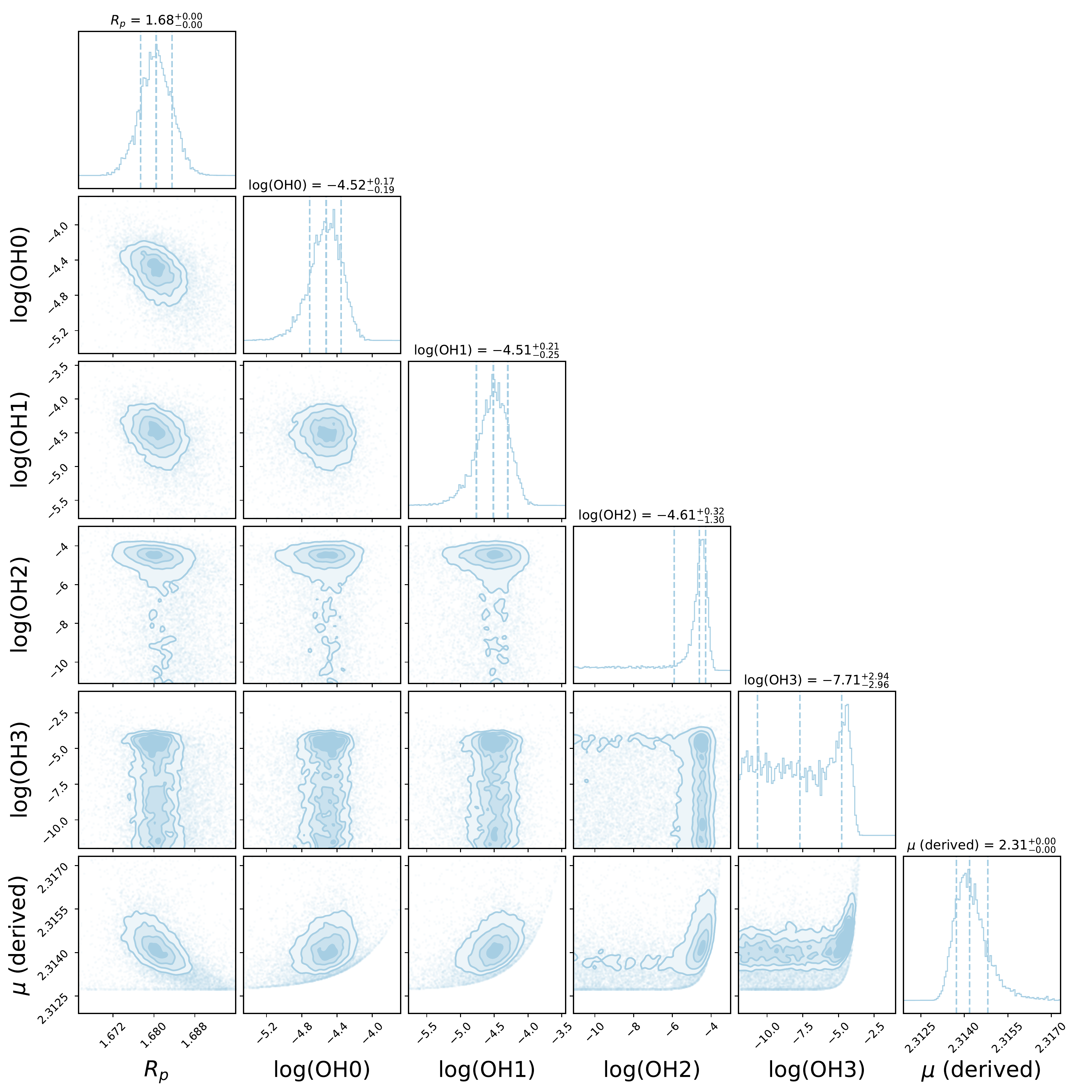}
    \caption{Retrieval nested sampling posteriors for one simulated secondary eclipse with bands 0 to 3.}
    \label{fig:0_3_1_constant_mix_posteriors}
\end{figure}

\begin{figure}[!ht]
    \centering
    \includegraphics[width=\columnwidth]{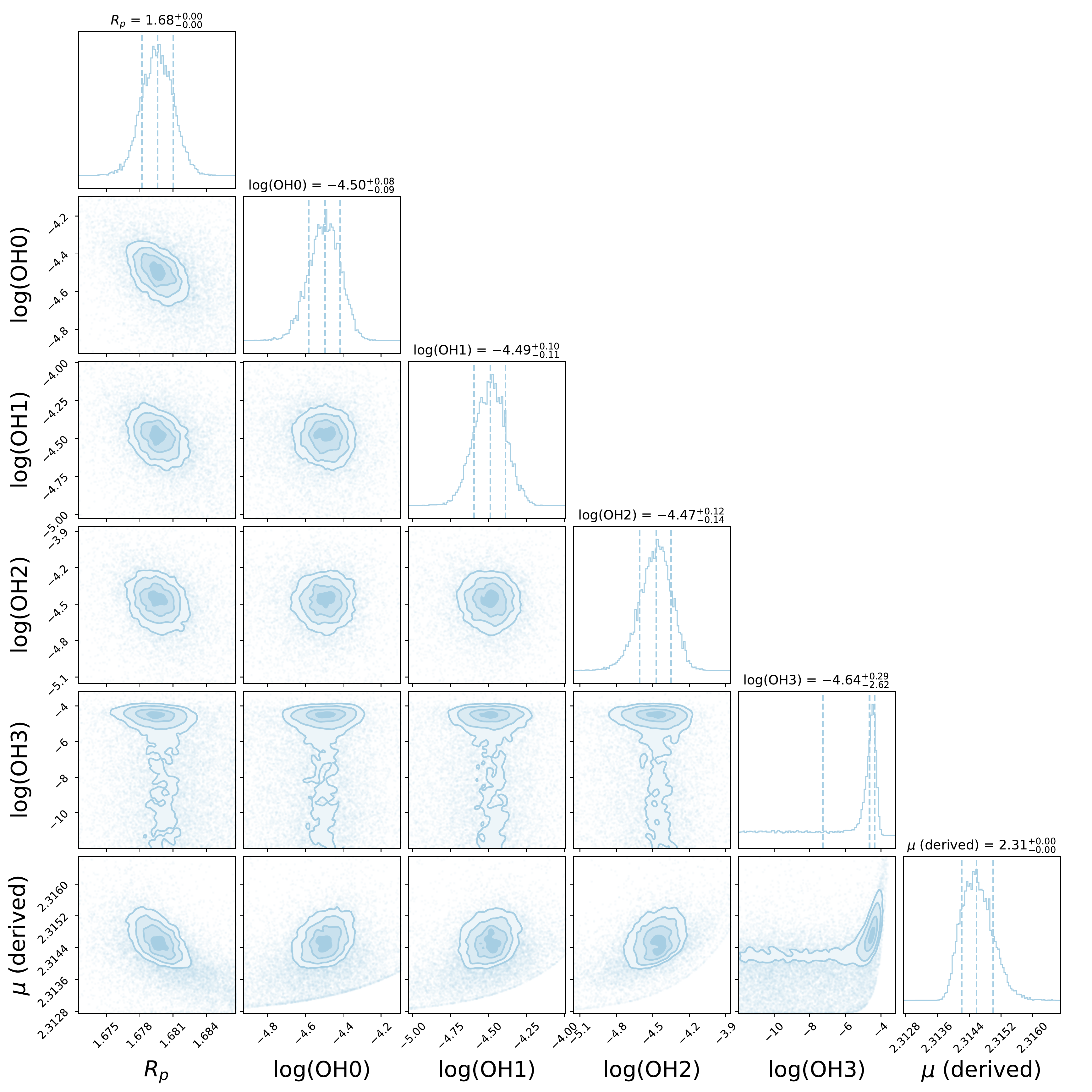}
    \caption{Retrieval nested sampling posteriors for five simulated secondary eclipses with bands 0 to 3.}
    \label{fig:0_3_5_constant_mix_posteriors}
\end{figure}

\begin{figure}[!ht]
    \centering
    \includegraphics[width=\columnwidth]{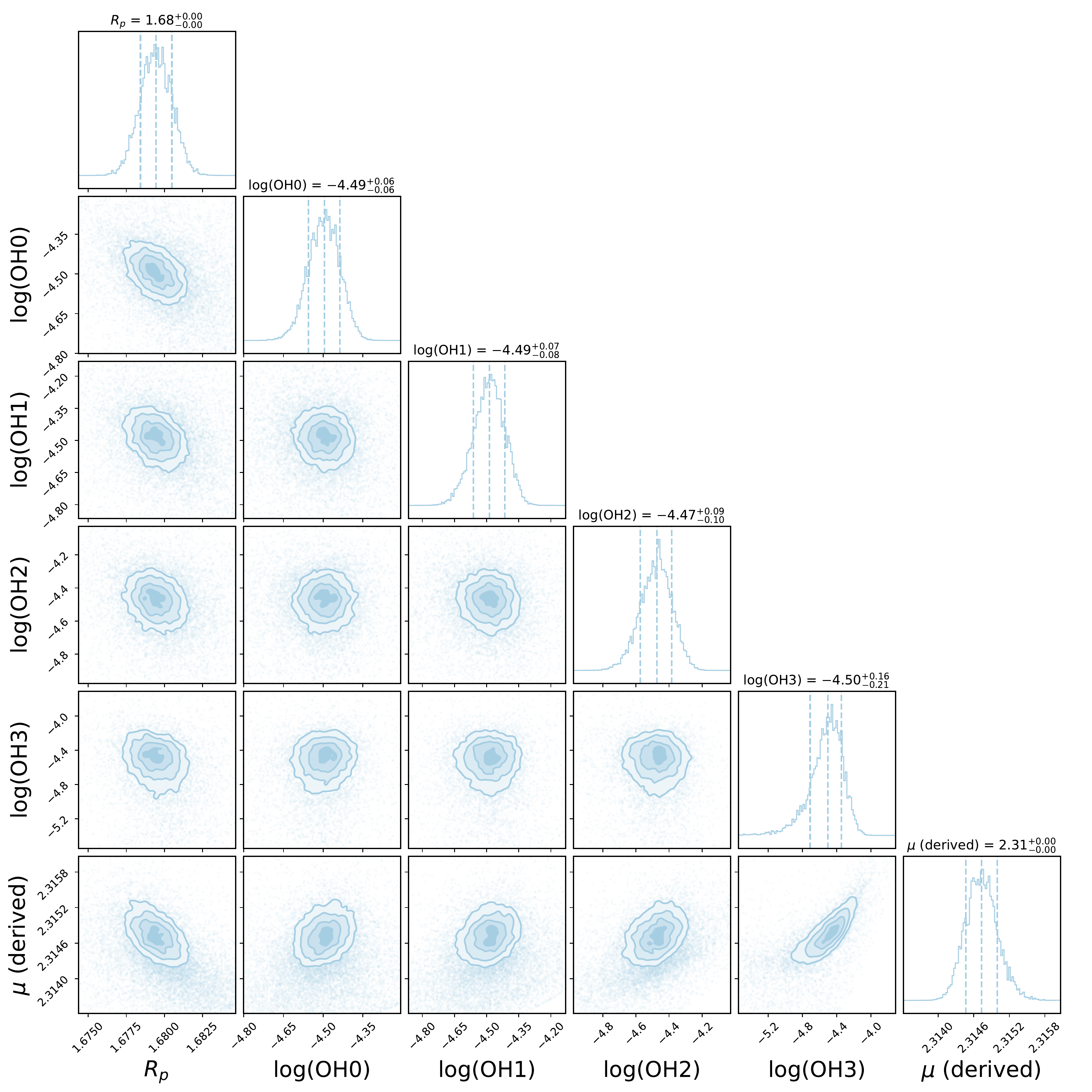}
    \caption{Retrieval nested sampling posteriors for ten simulated secondary eclipses with bands 0 to 3.}
    \label{fig:0_3_10_constant_mix_posteriors}
\end{figure}

Figure \ref{fig:0_2_1_constant_mix_posteriors} shows the retrieved posterior probability distributions with the three strongest emission bands of  OH in NIR with $\Delta v =2$ and $\Delta v= 3$, bands 0, 1, and 2 (i.e. corresponding to the emissions from three states $v'=2, 3$ and 4 with the highest populations). For  this three band case, treating each band as a separate species, a single secondary eclipse observation is sufficient to well constrain the individual band mixing ratios ($10^{-4.53}$, $10^{-4.51}$ and $10^{-4.60}$) to within reasonable agreement with the ground truth $10^{-4.2}$ (all within $2\sigma$).

When we include the hotter band 3, the lines of which are weaker on account of the reduced relative populations, additional secondary eclipses are required to well constrain the band mixing ratios. The posteriors for a single secondary eclipse retrieval are shown in Fig.~\ref{fig:0_3_1_constant_mix_posteriors}, while the posteriors for retrievals conducted on 5 and 10 secondary eclipses are shown in Figs.~\ref{fig:0_3_5_constant_mix_posteriors} and \ref{fig:0_3_10_constant_mix_posteriors} respectively. In this final retrieval, the band abundances are neatly constrained, with little degeneracy.

\subsection{Spectroscopic temperature fitting from vibrational populations}

We then attempted to deduce the temperature spectroscopically from the individual vibrational populations using the  simulated data, again treating each emission band individually. We used the  vibrational band cross sections with the population normalised to 1 ($N_v=1$) introduced in section~\ref{sec:xsecs}. The retrieved mixing ratio now acts as a proxy for the band population (in addition to the information it contains about the overall mixing ratio of OH in the atmospheric model). 

\begin{figure}[!ht]
    \centering
    \includegraphics[width=\columnwidth]{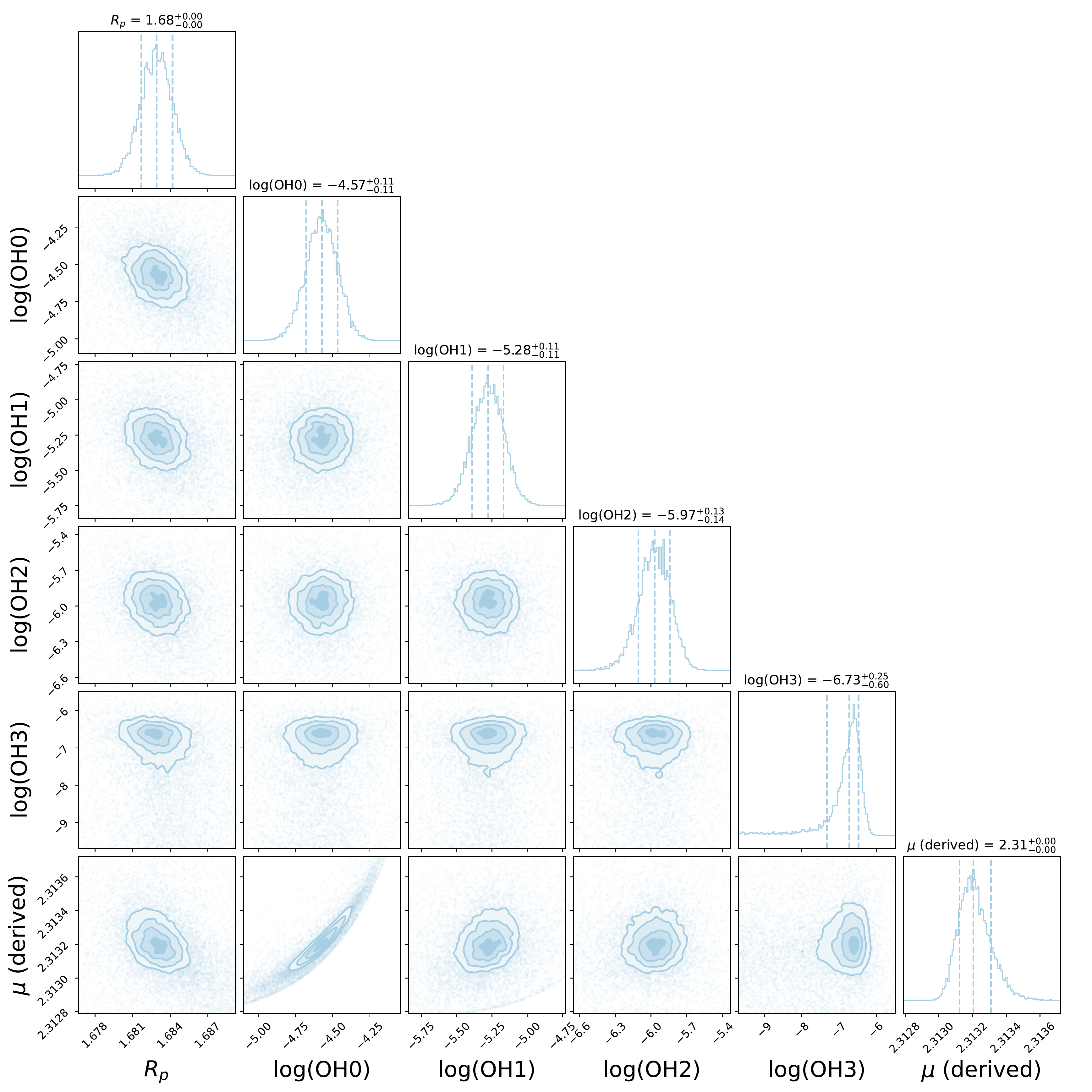}
    \caption{Posteriors for five simulated secondary eclipses with bands 0 to 3, using artificial completely populated band cross sections.}
    \label{fig:pop1d_0_3_5_constant_mix_posteriors}
\end{figure}

\begin{figure}[!ht]
    \centering
    \includegraphics[width=1.0\columnwidth]{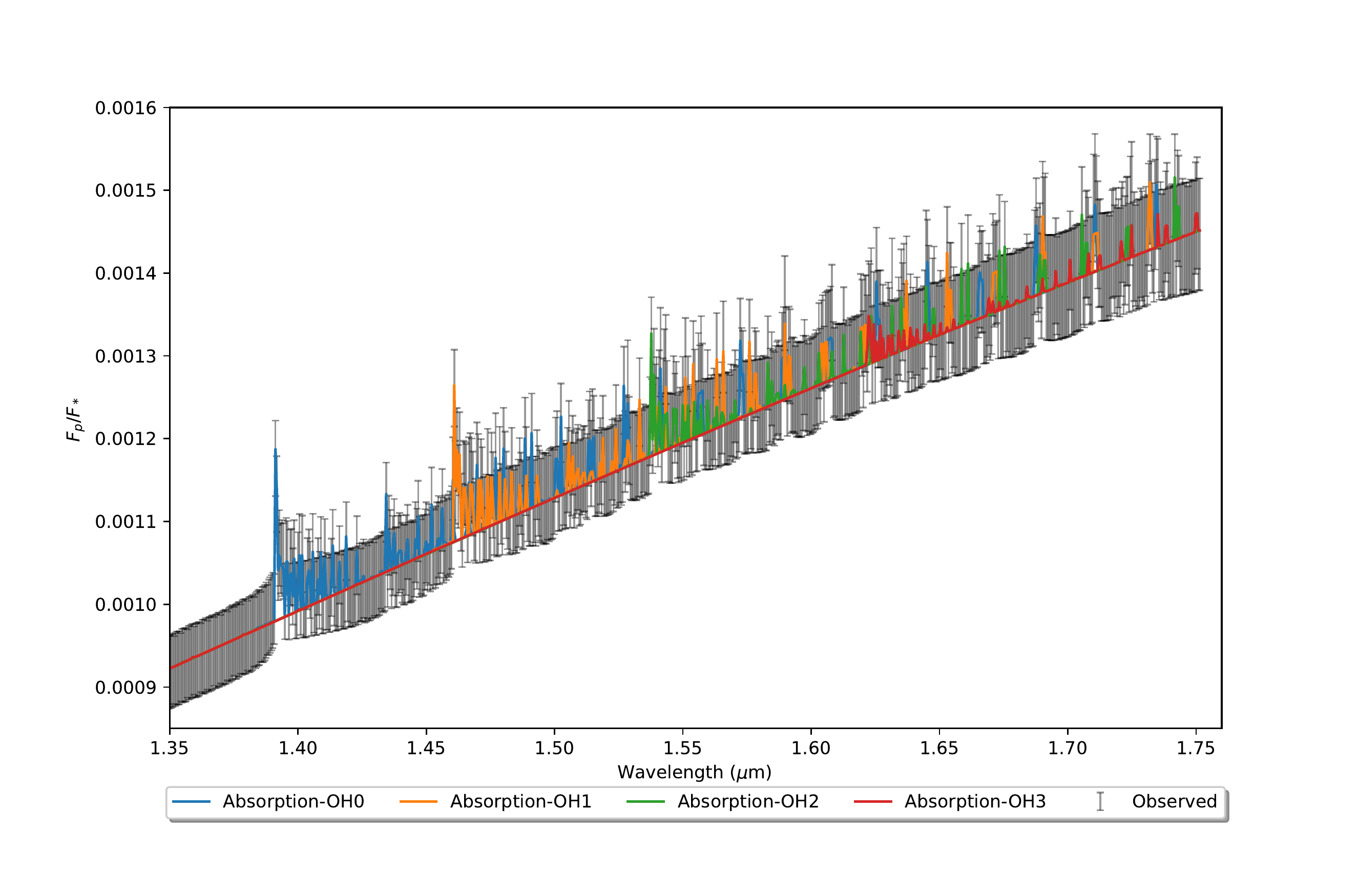}
    \caption{Individual band contributions of OH in NIR retrieved using the completely populated band cross sections for 0 to 3 with a simulated five secondary eclipses, with the plot highlighting the wavelength region with the strongest lines.}
    \label{fig:pop1d_0_3_5_constant_mix_contrib}
\end{figure}

\begin{figure*}[!ht]
    \centering
    \includegraphics[width=1.5\columnwidth]{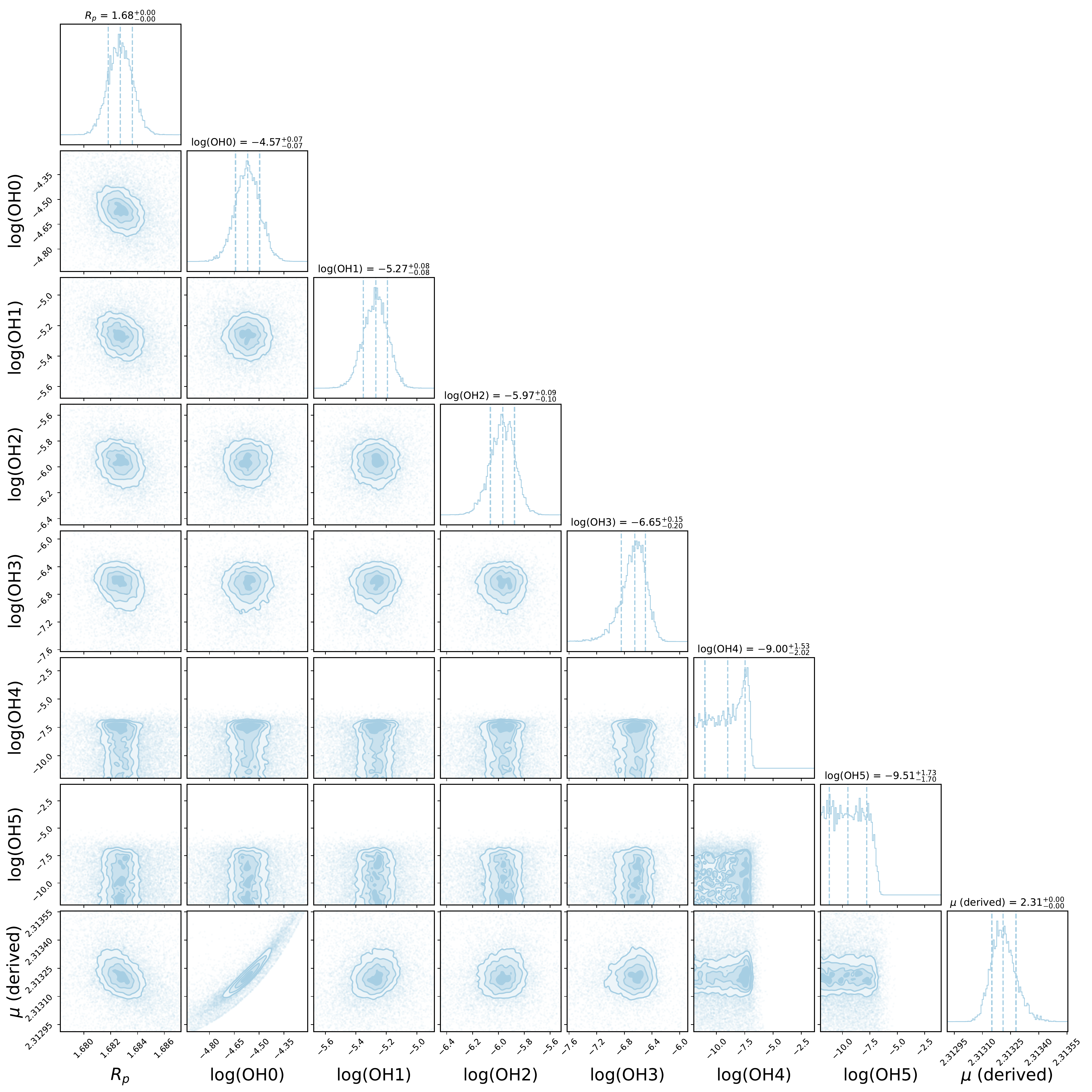}
    \caption{Posteriors for ten simulated secondary eclipses with bands 0 to 5, using artificial completely populated band cross sections.}
    \label{fig:pop1d_0_5_10_constant_mix_posteriors}
\end{figure*}

When considering the first three bands in isolation, using data simulated for five secondary eclipse observations, mixing ratios for each of the individual bands can be well constrained (see Fig.~\ref{fig:pop1d_0_3_5_constant_mix_posteriors}). The crucial difference compared to a typical retrieval of the mixing ratio for a molecule is that we see a drop off in retrieved mixing ratio with increasing $v''$, corresponding to the reduced populations in higher bands for OH in LTE at this temperature. The relative weakness of the lines due to these higher states can be visualised in Fig.~\ref{fig:pop1d_0_3_5_constant_mix_contrib} where the individual contributions used in the retrieval are plotted. The limit of our ability to retrieve such mixing ratios for less populated bands with a simulated ten secondary eclipses of data is apparent in Fig.~\ref{fig:pop1d_0_5_10_constant_mix_posteriors} and is well understood by their marginal contribution to the OH spectra as well as by the limitations of the instrument.

\begin{figure*}[!ht]
    \centering
    \includegraphics[width=1.5\columnwidth]{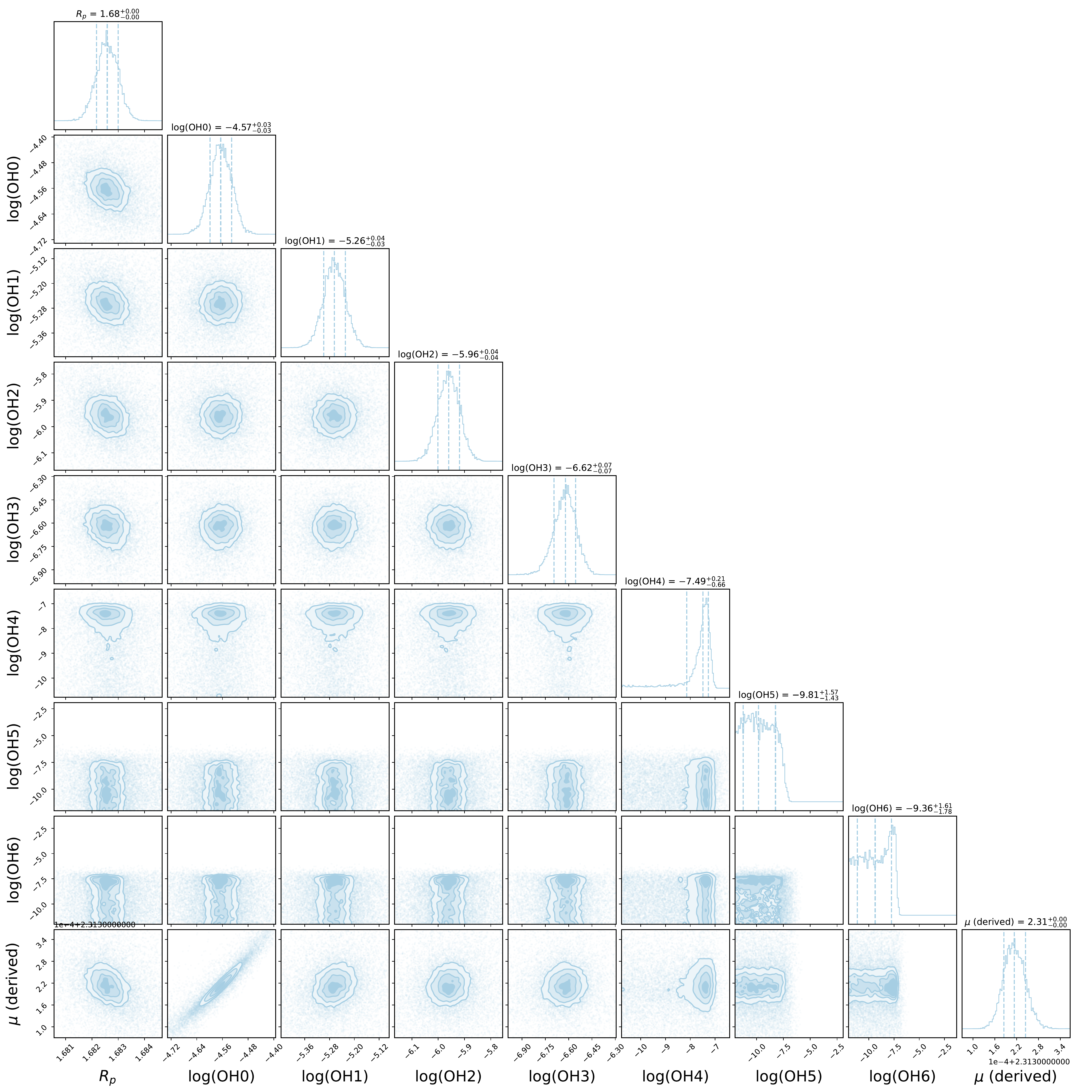}
    \caption{Posteriors for the retrieval including bands 0 through 6 for a simulated 50 secondary eclipses.}
    \label{fig:pop1d_0_6_50_constant_mix_posteriors}
\end{figure*}

\begin{figure}[!ht]
    \centering
    \includegraphics[width=\columnwidth]{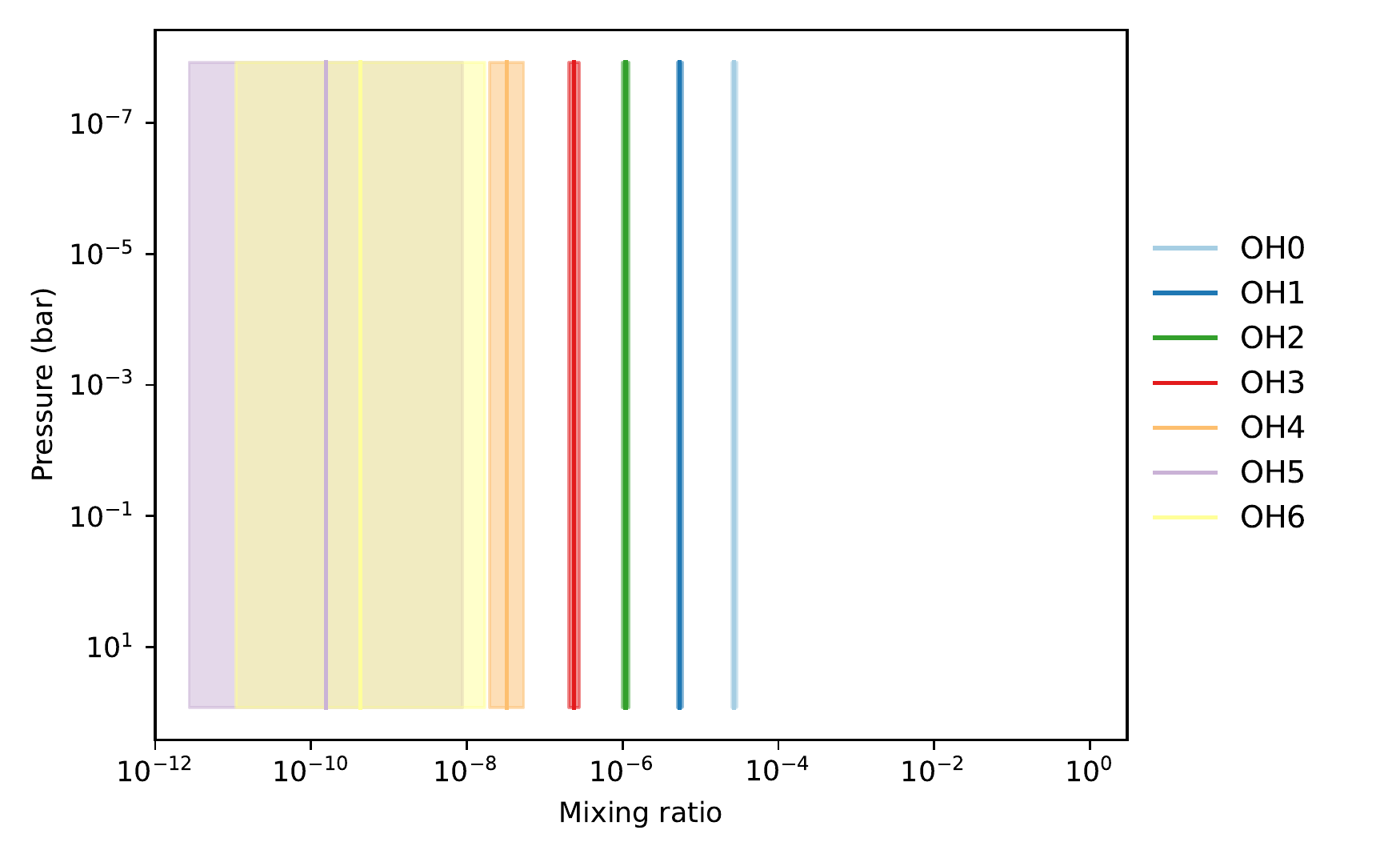}
    \caption{Mix ratios retrieved for bands 0 through 6  for a simulated 50 secondary eclipses.
    }
    \label{fig:pop1d_0_6_50_constant_mix_mixratio}
\end{figure}

In order to illustrate the limitations of detectability  imposed by the low state populations and the associated  limited emissions of  very hot bands,  in Fig.~\ref{fig:pop1d_0_6_50_constant_mix_posteriors} we examine the case of 50 secondary eclipse observations. One can see that even for this idealised case, only one more band can be constrained properly. For the subsequent two bands, only an upper limit can be given and the 1$\sigma$ confidence interval for the mixing ratio of these bands spans a wide, poorly constrained range, as shown in Fig.~\ref{fig:pop1d_0_6_50_constant_mix_mixratio}.

We can now use this 50 simulated secondary eclipse retrieval to demonstrate spectroscopic retrieval of the vibrational temperature $T_{\rm vib}$ via fitting the retrieved populations with a Boltzmann distribution. The band mixing ratios retrieved individually using the individual complete band sets were normalised by their sum. We then employ a simple least squares fitting routine with the vibrational temperature as our free parameter and the vibrational populations $N_v(T)$ are given by Eq.~(\ref{e:Nv}), from which we derive a best fit vibrational temperature of 2815~K. It should be noted that this simulated example is not an independent measurement of the thermal conditions, since the literature T-P profile used in simulation is assumed in the retrieval. Instead it demonstrates the retrievability of vibrational populations as a technique which could be used in conjunction with fitting the T-P profile when conducting such an analysis in the empirical case, with an unknown temperature profile.

\begin{figure}[!ht]
    \centering
    \includegraphics[width=\columnwidth]{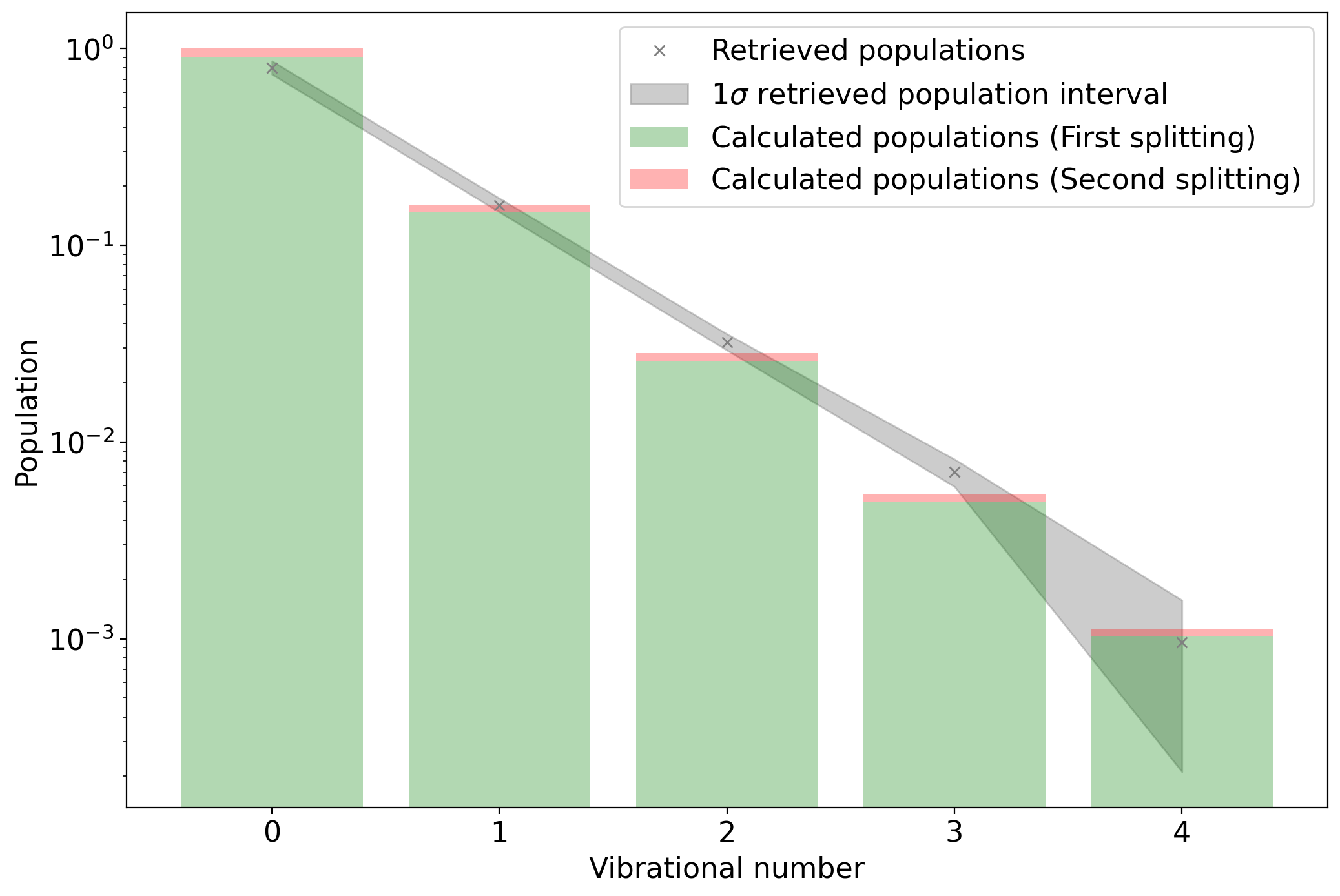}
    \caption{Retrieved mixing ratios for OH compared with the theoretical Boltzmann distribution used in generating cross sections for OH in LTE. Retrieved from a simulated 50 secondary eclipses.}
    \label{fig:retrieved_boltzmann}
\end{figure}

The agreement between the Boltzmann distribution and the retrieved mixing ratios for the bands, as proxies for the band populations, are plotted in Fig.~\ref{fig:retrieved_boltzmann}. The bars show two population levels for each vibrational number, with the values given in table~\ref{tab:pops}; these arise since OH is an open-shell molecule and so fine structure splitting gives rise to two energies per vibrational number.

\subsection{non-LTE retrieval}

\begin{figure}[!ht]
    \centering
    \includegraphics[width=\columnwidth]{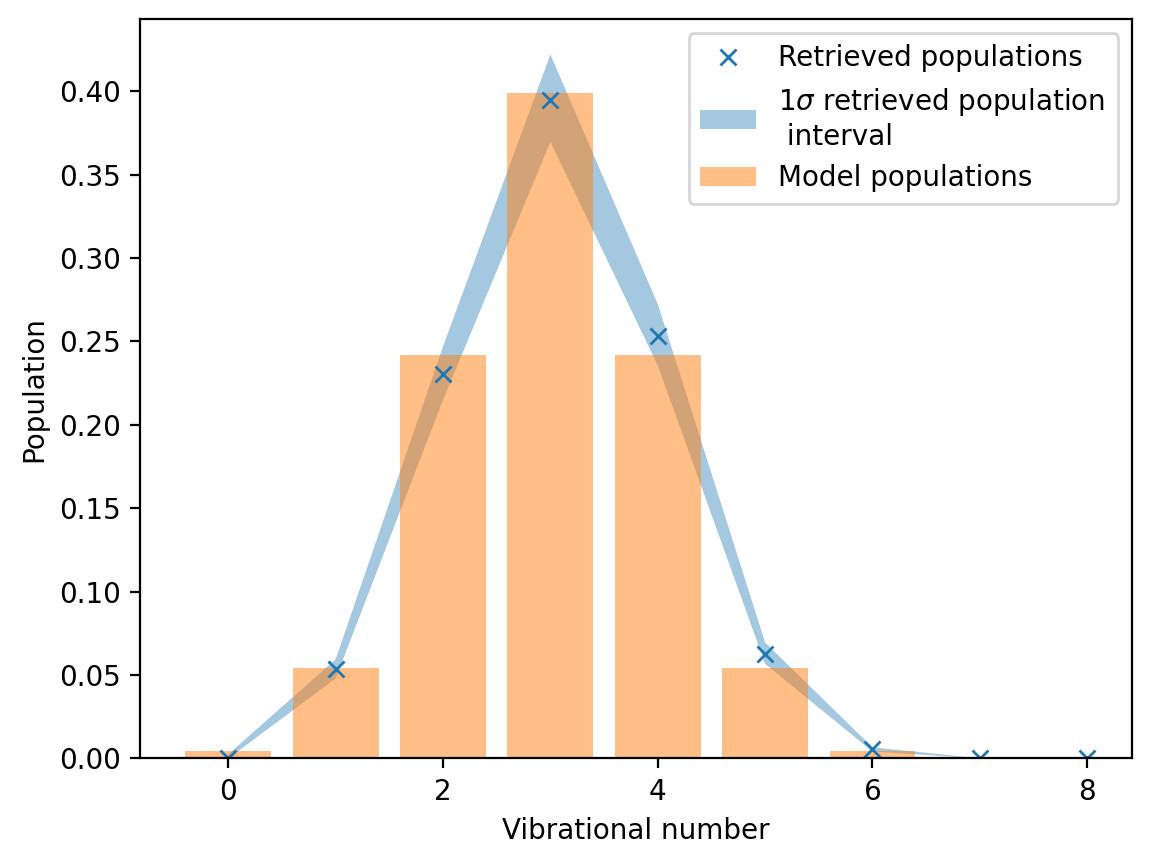}
    \caption{Retrieved mixing ratios for OH compared with a simulated spectra generated using OH with a disequilibrium, normal distribution peaking at $v=3$.}
    \label{fig:retrieved_normv3}
\end{figure}

In addition, we consider a hypothetical non-LTE scenario where the vibrational state populations have been driven away from a Boltzmann distribution, similar to that shown for the excited bending vibrational mode of H$_{2}$O by \citet{22HaZhCo.nLTE}.
In our non-LTE scenario, we take the molecules to have been vibrationally excited to an arbitrary peak of $v=3$ with the remaining state populations following a normal distribution around this peak, as shown in table~\ref{tab:pops}. This normally distributed configuration is an example of non-Boltzmann population; the retrieval result is illustrated in Fig.~\ref{fig:retrieved_normv3}.  An example of a physically occurring non-LTE scenario is the non-LTE vibrational population of OH in the air glow in the upper layers of the terrestrial atmosphere \citep{21ChHuGu.OH,20NoWiGo.OH}. 

\begin{table*}
\centering
\begin{tabular}{l|l|l|l}
$v$ & Normal population & Boltzmann population spin 1  & Boltzmann population spin 2\\
\hline
\hline
0 & 0.0044 & 1.0000 & 0.9086\\
1 & 0.0540 & 0.1614 & 0.1467\\
2 & 0.2420 & 0.0283 & 0.0258\\
3 & 0.3990 & 0.0054 & 0.0049\\
4 & 0.2420 & 0.0011 & 0.0010\\
5 & 0.0540 & - & -\\
6 & 0.0044 & - & -\\
\hline
\end{tabular}
\caption{Populations corresponding to $v$ states for the normal distribution (a non-LTE) example and the Boltzmann LTE example for both populations arising from the spin splitting. \label{tab:pops}
}
\end{table*}

\section{High resolution cross-correlation approach}

A high resolution cross-correlation approach comprises part of this analysis. While we have demonstrated the theoretical capability to separate individual bands at the modelling stage and to integrate this into a retrieval framework, we were motivated by the limited resolution of such data from space-based observatories to investigate the use of high resolution data from ground-based facilities. In this case, that is the IRD instrument on the Subaru telescope. To achieve this we perform high resolution cross-correlation fits for the free parameters: orbital velocity $K_{\mathrm{P}}$, systemic velocity $v_{\mathrm{sys}}$, and the scale factor of the model $\alpha$. These fits are performed for a number of forward models, each of which has a fixed set of vibrational band contributions.

\subsection{Observations and Data Reduction}
We used three data sets taken after the secondary eclipse using IRD mounted at the Subaru 8.2-m telescope. The first data set was obtained on October 1, 2020 (N1, PID: S20B-008, PI: Nugroho) and used to detect OH emission, see \citet{21NuKaHa.OH}. While the additional two data sets were taken on August 30, 2021 (N2) and December 18, 2021 (N3) as part of program S21B-111 (PI: Nugroho, see Table \ref{Tab2} for the observation log). The target was continuously observed in natural guide star mode without the laser frequency comb with an exposure time of 300\, s per frame and overhead time of 25 s. We discarded the first nine exposures of N3 which had low S/N due to high humidity. These data sets are part of the work of Nugroho et al. (in prep.) and we refer the reader there for full details.

\begin{table*}

\begin{tabular}{c|c|c|c|c|c|c|c}
\hline
Name of & Date & Observing time & Airmass & Phase & Number & Mean S/N& \# {\sc SysRem}\\
the night & (UT) & (UT) & change & coverage & of frames &(at 1 $\mu$m) & iterations\\
\hline
\hline
N1 & 2020-10-01 & 10:35-13:42 & 1.12-1.05-1.12 & 0.597-0.700 & 33& 145 & 3\\
N2 & 2021-08-30 & 10:25-15:15 & 1.64-1.05-1.08 & 0.569-0.734 & 56& 156 & 9\\
N3 & 2021-12-18 & 05:24-10:18 & 1.13-1.05-1.44 & 0.573-0.740 & 55& 110 & 6\\ 
\hline
\end{tabular}
\caption{Summary of IRD observations \label{Tab2}}
\end{table*}

In short, the data were reduced using the IRD pipeline \citep[][Kuzuhara et al. in prep.]{Kuzuhara2018}. The extracted spectra consist of 72 spectral orders ranging from $\approx$9260 to $\approx$17645 \AA\,. We exclude the first eight and the last spectral orders due to relatively low S/N. In addition, we also exclude the spectral orders with the median wavelength between 14600-15203 \AA\ from further analysis due to heavy telluric contamination. We fit the continuum profile of the spectrum with the highest S/N of each data-set using the {\sc continuum} task in {\sc IRAF}\footnote{The Image Reduction and Analysis Facility ({\sc IRAF}) is distributed by the US National Optical Astronomy Observatories, operated by the Association of Universities for Research in Astronomy, Inc., under a cooperative agreement with the National Science Foundation.}. Then we divided the data sets by their corresponding best-fit continuum profile to normalize them. The sky emission lines were masked based on \citet{Oliva2015, Rousselot2000}. The spectra were put into a "common blaze function/continuum profile" following the procedure of \citet{Merrit2020}. Any 5~$\sigma$ outliers were masked from the data following \citet{gibson2020}. Any remaining bad pixels and bad regions were visually identified and masked out from the data. The final spectra of each data set were then stacked into an array with three axes: number of spectral orders (58), number of exposures, and wavelength (2048). Finally, we also masked the pixels that have a value less than 0.1 from the data to avoid any saturated telluric lines and estimated the uncertainty of each pixel by calculating the outer product of the standard deviation along the wavelength and time, then normalized by the standard deviation of the whole array excluding the masked pixels.

In the wavelength range of IRD, telluric and stellar lines dominate the data. The signal of the planet is several orders of magnitude weaker than the signal of the host star. To extract the planet signal using the cross-correlation technique pioneered by \citet{Snellen2010}, the telluric and stellar lines must be removed from the data. Fortunately, the radial velocity of the planet ranged from tens to hundreds of km\,s$^{-1}$, shifting the planet spectra across many different pixels, while the telluric and stellar lines were (quasi-) stationary during the observations. To remove them, first, we divided out the mean spectra from each data set. We then applied the detrending algorithm {\sc SysRem} \citep{Tamuz2005} independently to each spectral order in each data set to remove the temporal variations in each wavelength bin. This was first applied to high-resolution datasets by \citet{Birkby2013}. Any wavelength bin in the residual of each iteration of each spectral order that has a standard deviation of more than 5 times the standard deviation of the whole array was masked. We stopped the iteration when the detection strength of OH is at its maximum (see Table \ref{Tab2} for the number of iterations selected for each data set).

\subsection{Cross-correlation and Likelihood Mapping}
We cross-correlated the telluric-cleaned data sets with our forward models Doppler-shifted using linear interpolation over a range of radial velocity. We used the cross-correlation function of \citet{gibson2020}:
\begin{equation}
    \mathrm{CCF}(v)= \sum_{i} \frac{f_{i}m_{i}(\nu)}{\sigma_{i}^{2}},
\end{equation}
where CCF is the cross-correlation function for a given Doppler-shifted velocity ($\nu$), $f_{i}$ is the mean-subtracted telluric-cleaned data at $i$th wavelength bin, $\sigma_{i}$ is the uncertainty, and $m_{i}$ is the mean-subtracted model Doppler-shifted to a radial velocity of $\nu$. 

The cross-correlation functions of all of the data sets were then co-added at the planetary rest frame over a range of orbital velocity ($K_{\mathrm{P}}$) and systemic velocity ($v_{\mathrm{sys}}$) creating a $K_{\mathrm{P}}$-$v_{\mathrm{sys}}$ CCF map. We then converted it to an S/N map by dividing it by the standard deviation of the area $\pm$ 50 km\,s$^{-1}$ away from the expected signal. Based on the previous studies, the planetary signal, if any, would appear around a $K_{\mathrm{P}}$ of 230\,km s$^{-1}$ and $v_{\mathrm{sys}}$ of -3\,km s$^{-1}$ \citep[e.g.,][]{Nugroho2017, Nugroho2020ApJL, Nugroho2021, Cont2021, Cont2022, Yan2021, vanSluijs2022, Yan2022b}. 

To avoid any bias in selecting which area to represent the noise, we followed \citet{Brogi2022} by fitting a Gaussian function to the CCF distribution of the noise and used the best-fit standard deviation as the estimated noise floor.

We also calculated a likelihood map ($\mathcal{L}$) for each model using the $\beta$-optimised likelihood function following \citet[][see also \citealt{19BrLixx}]{gibson2020}.
\begin{equation}\label{eq:lnlikelihood}
    \ln \mathcal{L}(v)= -\frac{N}{2} \ln \left[\frac{1}{N} \left( \sum\frac{f_{i}^{2}}{\sigma_{i}^{2}} + \alpha^{2}\sum \frac{m_{i}(\nu)^{2}}{\sigma_{i}^{2}}-2\alpha \mathrm{CCF(\nu)}\right) \right],
\end{equation}
where $\alpha$ is the scale factor of the model and $N$ is the total number of pixels. 

Previous studies \citep[e.g.][]{19BrLixx, Gibson2022} show that removing the telluric lines using a detrending method (e.g. airmass detrending, singular value decomposition, {\sc SysRem}) alters the planetary spectrum in the data. As cross-correlation or likelihood mapping is basically a template matching if the model is processed differently than the observed data any constraint that relies on the line profile (strength, position, width, and shape) would be unreliable. To correct this effect, for each spectral order of each data set, we Doppler-shifted the model template for a given value of $K_{\mathrm{P}}$ and $v_{\mathrm{sys}}$ as a function of orbital phase and divided out the mean spectrum to normalize it. We then applied the preprocessing technique described by \citet{Gibson2022} that uses the same {\sc SysRem} basis vectors that were used to remove the telluric lines and subtracted them out from the normalized Doppler-shifted model array. This preprocessed model array was then directly used to calculate the likelihood value. We repeated this process for a range of $K_{\mathrm{P}}$ and $v_{\mathrm{sys}}$. To account for different scaling of the model to the data, we calculated the likelihood for a range of $\alpha$ values. Lastly, previous literature \citep[e.g., ][]{gibson2020, Nugroho2020ApJL, Gibson2022} estimated the significance of the detection by dividing the median values of the conditional distribution of $\alpha$ at the best-fit parameters by its standard deviation. Here, we expanded it to produce a significance map by performing it to all of the combinations of $K_{\mathrm{P}}$ and $v_{\mathrm{sys}}$.

\subsection{LTE forward model-all bands}
\begin{figure}[!ht]
    \centering
    \includegraphics[width=\columnwidth]{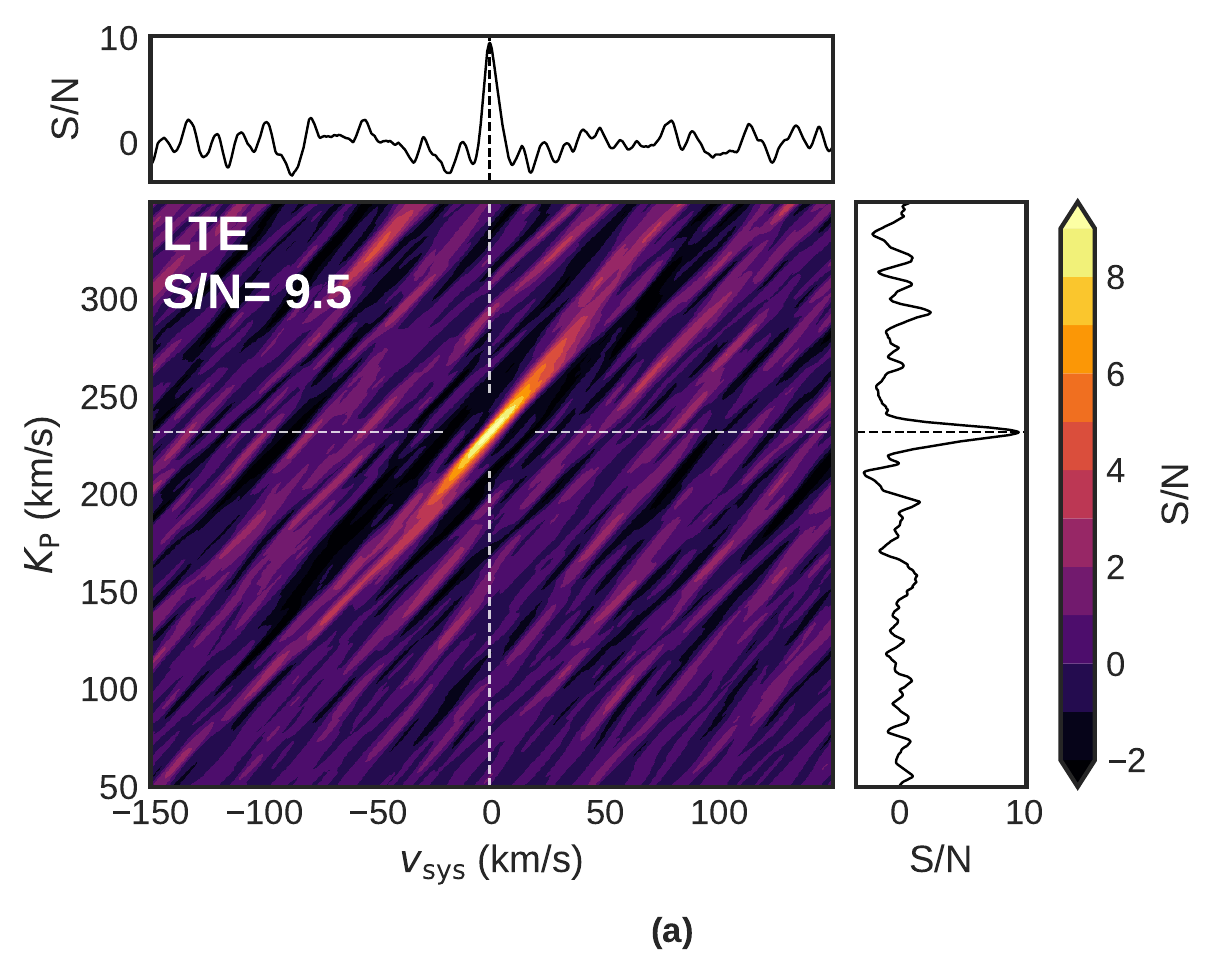}
    \includegraphics[width=\columnwidth]{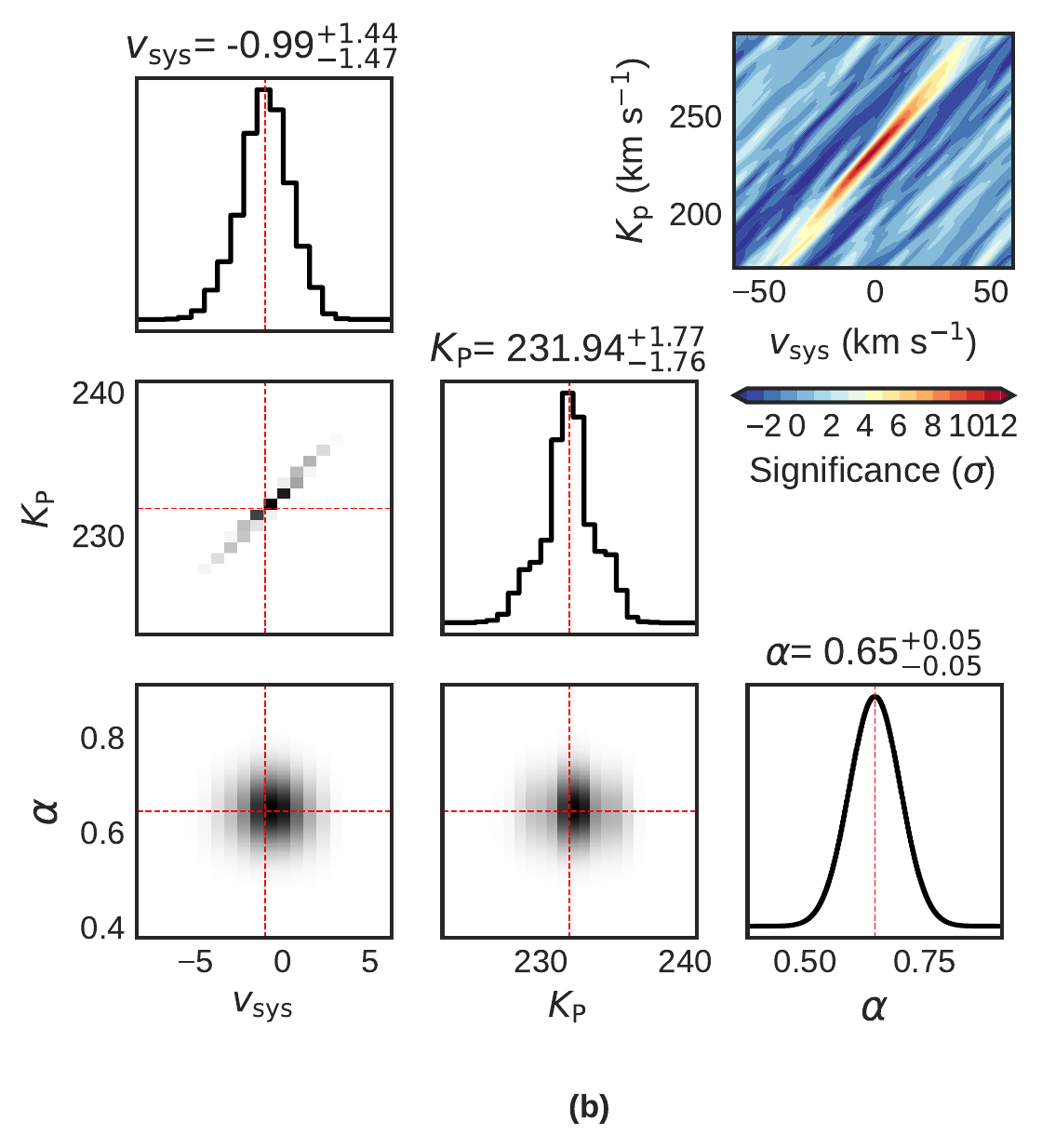}
    \caption{(a). The $K_{\mathrm{P}}$-$v_{\mathrm{sys}}$ map after combining the cross-correlation functions of all of the data sets using a non-preprocessed OH LTE model. The upper panel shows the S/N curve at the $K_{\mathrm{P}}$ of 231.8 \,km s$^{-1}$ while the right panel shows the S/N curve at the $v_{\mathrm{sys}}$ of -1.1\,km s$^{-1}$. The color-bar shows the S/N of the $K_{\mathrm{P}}$-$v_{\mathrm{sys}}$ map.
    (b). The marginalized likelihood distribution of $K_{\mathrm{P}}$, $v_{\mathrm{sys}}$, and $\alpha$ using a preprocessed OH LTE model. The red dashed lines show the median value of the corresponding distribution. The upper right panel is the $K_{\mathrm{P}}$-$v_{\mathrm{sys}}$ map of the detection from the likelihood. The color-bar shows the significance of the $K_{\mathrm{P}}$-$v_{\mathrm{sys}}$ map. This map is produced by dividing the median value of the conditional likelihood distribution of $\alpha$ by its uncertainty at every combination of $K_{\mathrm{P}}$ and $v_{\mathrm{sys}}$.}
    \label{fig:kpvsys_OH_LTE_allband}
\end{figure}

Using our OH LTE forward model, we obtained a strong detection of OH emission with an S/N of 9.5 at the $K_{\mathrm{P}}$ of 231.8 \,km s$^{-1}$ and $v_{\mathrm{sys}}$ of $-$1.1\,km s$^{-1}$ consistent with the previous detection (Fig. \ref{fig:kpvsys_OH_LTE_allband}a). With the preprocessed model and likelihood mapping, the same signal is detected at $>$ 12 $\sigma$ (Fig. \ref{fig:kpvsys_OH_LTE_allband}b) demonstrating the advantage of the likelihood framework and preprocessing the model over a pure cross-correlation without preprocessing the model in detecting atmospheric signal.

\subsection{Individual bands} \label{subsection:indv_band}
\begin{figure*}[!ht]
    \centering
    \includegraphics[width=0.85\textwidth]{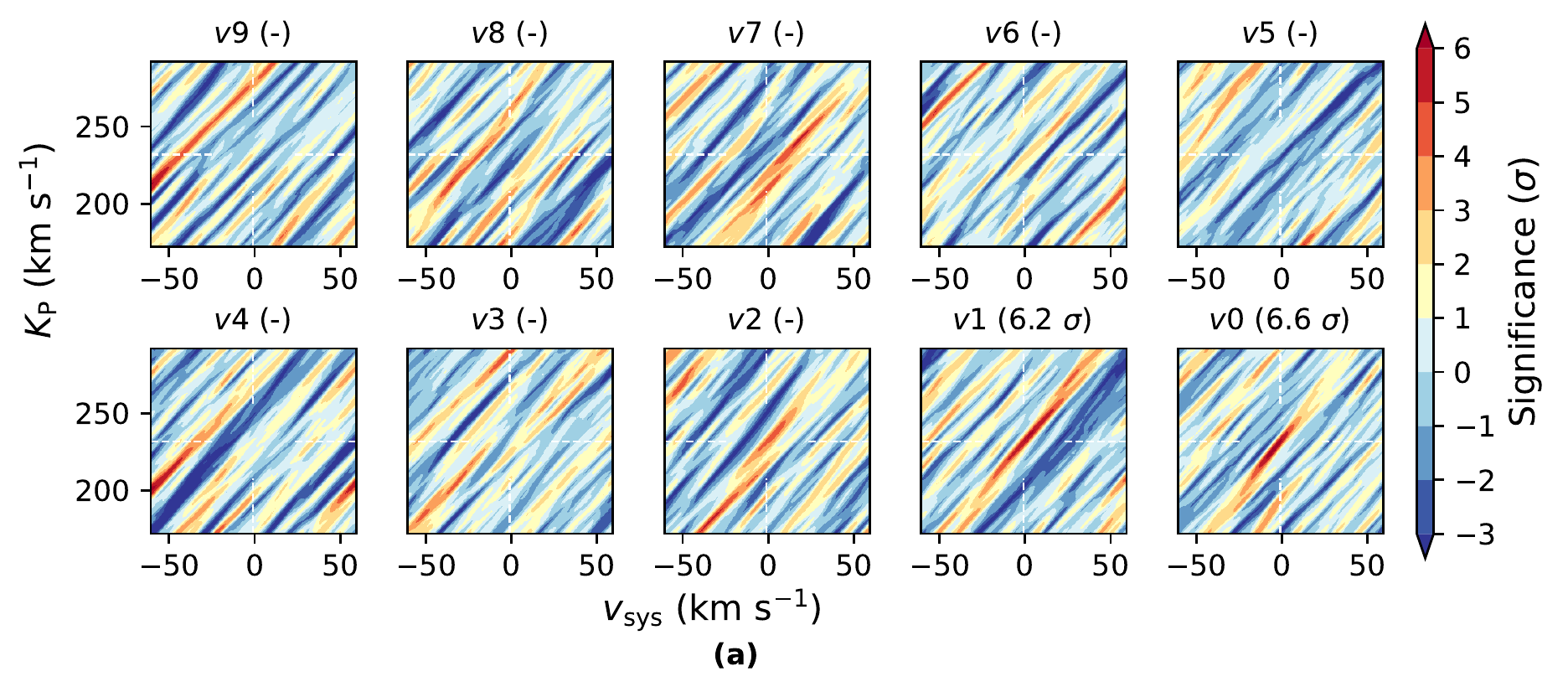}
    \includegraphics[width=0.9\textwidth]{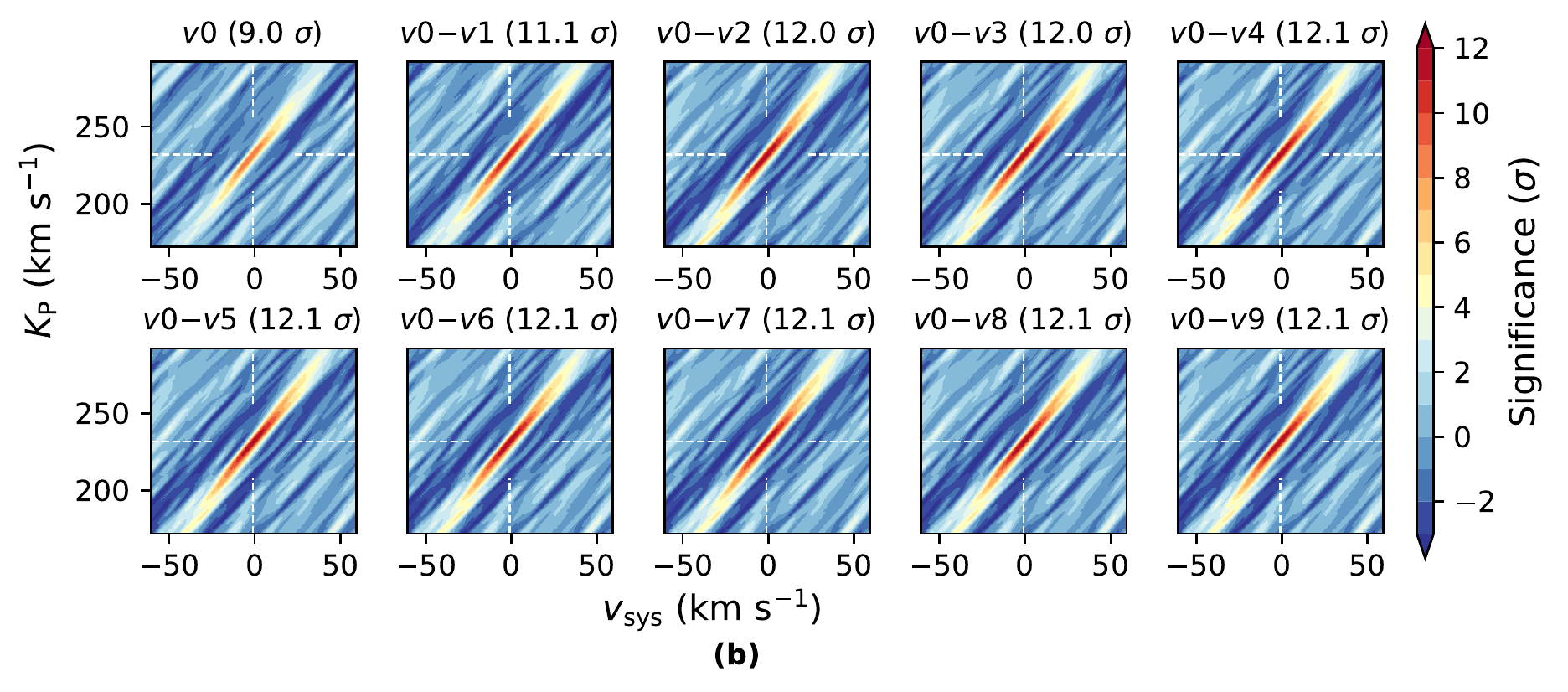}
    \includegraphics[width=0.9\textwidth]{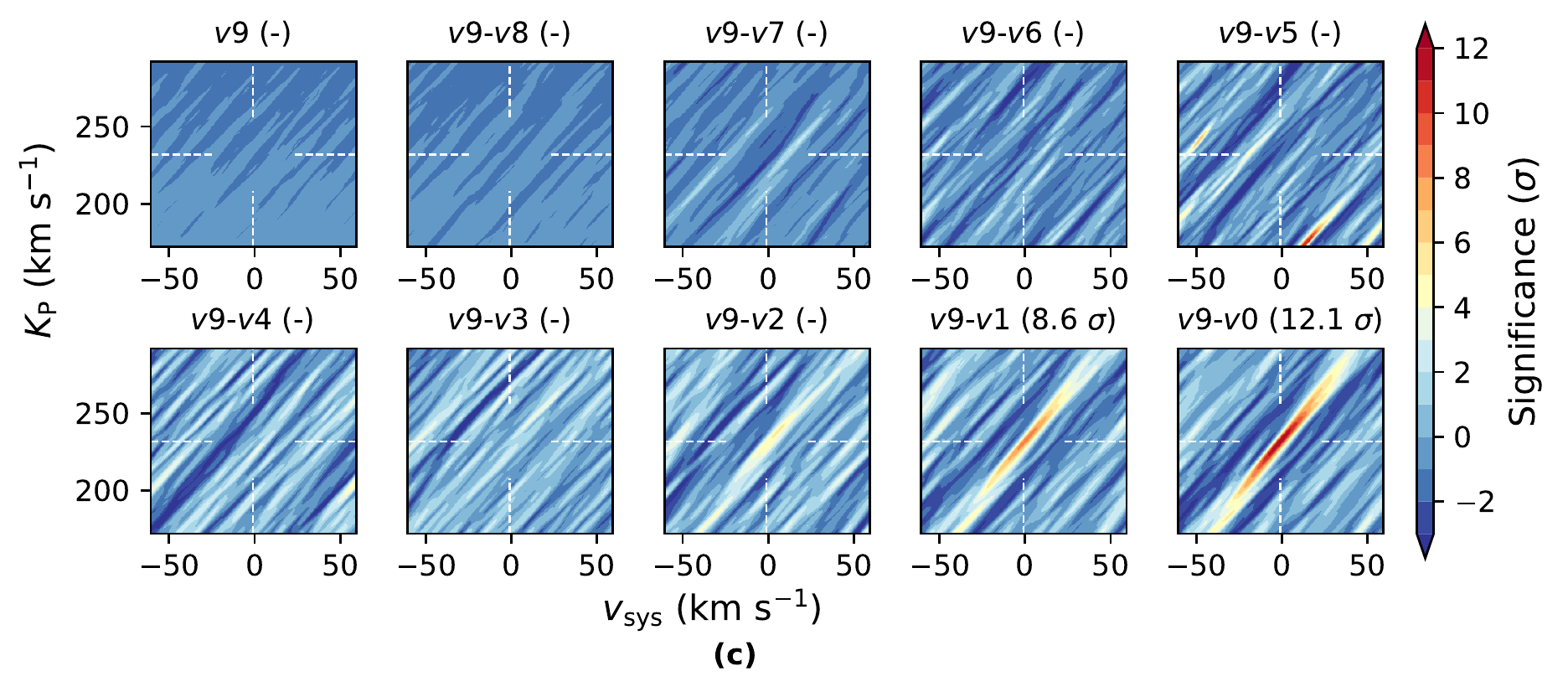}
    \caption{The $K_{\mathrm{P}}$-$v_{\mathrm{sys}}$ significance map of the combined data sets ($N_1+N_2+N_3$) for the preprocessed individual band model (\textbf{a}), unpeeling model from band 0 only to the total 0+1+2+3+4+5+6+7+8+9 bands (\textbf{b}), unpeeling model from band 9 only to the sum of bands 0, 1, 2, 3, 4, 5, 6, 7, 8, and 9 (\textbf{c}). The white dashed lines indicate the location of the OH LTE model signal as a reference. The color bar shows the significance of the maps.}
    \label{fig:kpvsys_OH_unpeeling}
\end{figure*}

Here, we tried to detect the individual OH band assuming a completely populated state for each band similar to Fig.~\ref{fig:pop1d_xsecs} Rotational state populations are taken to be Boltzmann distributed, while the vibrational state populations are unweighted in order to explore any non-LTE (non-Boltzmann) circumstances. For the high resolution case we extended our isolated cross section set up to $v=9$ (which corresponds to a maximum highest state of $v = 12$ via the $\Delta v = 3$ band) in order to reflect the potential for higher sensitivty than the lower resolution modelling. We obtained a strong detection both for bands 0 and 1 at $>$ 6 $\sigma$ consistent with the velocity of the planet detected using the OH LTE model. Unfortunately, there is no clear evidence for hotter vibrational bands, although there is a hint of a weak signal of band 2 in the $K_{\mathrm{P}}$-$v_{\mathrm{sys}}$ significance map (see Fig. \ref{fig:kpvsys_OH_unpeeling}a). 

\subsection{Band unpeeling}\label{subsection:unpeeling}

High-resolution spectroscopy, obtained from ground-based instruments, contains a substantial noise component due to sources such as the telluric lines from the Earth's atmosphere; in comparison the component from the actual Exoplanet's atmosphere has a low signal-to-noise ratio. With this limitation, and since the individual bands in isolation provide even smaller signal contributions, we obtain inconclusive results when performing the cross-correlation technique with templates containing isolated vibrational bands. 

To mitigate the limited contributions from individual bands we attempt a spectral `unpeeling' approach; cross-correlation templates were constructed with bands sequentially added from 0 to 9 (i.e. bands 0, 0+1, 0+1+2, 0+1+2+3, etc), and also backwards from 9 to 0 (i.e. bands 9, 9+8, 9+8+7, \ldots, 9+8+7+6+5+4+3+2+1+0). We propose the unpeeling approach to include contributions from a greater number of bands in the model templates. This is to present an approach which may enhance cross-correlation signal in circumstances where individual band signals are too weak, while still performing some band isolation across sequential models as is the goal of this technique. The combination of a number of the vibrational bands then yielded a recognizable change in the detection significance when comparing cross-correlation results between different templates (see Fig. \ref{fig:kpvsys_OH_unpeeling}b \& \ref{fig:kpvsys_OH_unpeeling}c). This change is especially significant from 0 to 0+1 consistent with the 0 and 1 states being mainly populated at this temperature (see Section \ref{subsection:indv_band} and Fig. \ref{fig:kpvsys_OH_unpeeling}a) and the high detection significance of backwards unpeeling model 9+8+7+6+5+4+3+2+1 (i.e. with band 0 excluded). Meanwhile, the  9+8+7+6+5+4+3+2 backward model (with bands 0 and 1 excluded) does not yield any detection although visually we can see a possible signal of this model around the expected velocity of the planet similar to the individual band 2  model in Section \ref{subsection:indv_band}. This is also supported by the increase of the detection significance when  band 2 is added into the 0+1 band model. Interestingly, the single band 0 of the unpeeling band model yields a much stronger detection (9$\sigma$) than the completely populated (unity scaled) $v0$ case (6.6$\sigma$). This difference in the detection significances arises from a difference in the model templates, driven by differing line intensities. These line intensities are governed by the vibrational state populations which in the case of the completely populated model have been artificially set to unity, while in the filtered case the populations obey a Boltzmann distribution but with filtering applied afterwards to select the $v0$ model.  

The difficulty of detecting vibrational hot bands is in principle in accordance with the LTE conditions corresponding to the T-P profile temperature; the vibrational state populations relate to the detection significances for each individual band model, with only the first two detected in line with Boltzmann distributed populations corresponding to a vibrational temperature consistent with the effective temperature. This difficulty of detection is also related to the limited NIR region used, which does not properly cover the contributions of the hot bands  $v''>5$.

\label{sec:cross_correlation}

\section{Discussion} \label{sec:discussion}

Although the wavelength separation is good for the individual hot bands of molecules such as OH, the ability to distinguish between the bands in all situations is not unlimited and sufficient wavelength range coverage is required to avoid degeneracy. This limitation becomes apparent when attempting retrievals on spectra for state populations peaking at higher vibrational numbers.  For instance, using the current wavelength coverage of 0.97-1.5~\um\ in  atmospheric retrievals, it would be difficult to detect the terrestrial-like  airglow population, which is known to peak at the higher vibrational band  $(8,6)$ \citep{21ChHuGu.OH}. When attempting a retrieval of this non-LTE scenario, we encountered a substantial overestimation of the corresponding  vibrational band, $(7,5)$ or $(8,6)$, abundances characterised by a convergence to the top boundary of the prior when fitting via a nested sampling method. This problem exists for these higher bands when observing over this wavelength range since the stronger band becomes truncated, to constrain these bands a wavelength coverage reaching higher wavelengths would be required.

As the ground truth population for a given state becomes small, the associated transition lines in the spectrum become weaker. When this weakness becomes sufficiently small compared to the observation uncertainty (simulated, in this case), the populations for these states are no longer retrievable. This can be seen in Fig.~\ref{fig:retrieved_boltzmann}, where it is characterised by a far wider $1\sigma$ interval for the retrieved $v=4$ population.

It is theoretically possible to add an additional level of complexity and also attempt retrieval of fine structure spin splitting state populations, for which theoretical values are represented by the overlapping bars for each vibrational number in Fig.~\ref{fig:retrieved_boltzmann}. OH in this electronic configuration is characterised by spin splitting of 126.3\cm. The quantum number assignments available in this line list are sufficient to do this theoretically, however since observational data quality is limited, we leave this for future work.

WASP-33 is a pulsating $\delta$-Scuti variable \citep{22ChEdAl.exo} which makes it a slightly more complex system; this should not be an issue for the high resolution analysis as there is no OH in the stellar atmosphere to affect the cross-correlation results.

Possible further work could explore using this technique for a range of other molecules (such as TiO or other diatomics due to their good vibrational band separation) and a range of other data sets; including different wavelength ranges and target planets. One such candidate could be WASP-18b where OH has also been detected \citep{Brogi2022}.

\section{Conclusion} \label{sec:conclusions}

From the analysis conducted here, we can conclude that individual vibrational state populations can be obtained for OH in NIR  as fitted parameters in exoplanet atmosphere retrieval techniques by including isolated individual bands as cross sections of individual species. 
Here we have shown the viability of this approach for OH, though this generalises to diatomic molecules more broadly on account of their good vibrational band separation. Not only is this shown for simulated observational spectra for OH with a typical LTE Boltzmann distribution across vibrational states, but also for a non equilibrium case. In addition, we demonstrate that non Boltzmann distributions are also retrievable: for the non equilibrium case, this approach can also distinguish where higher states are populated, as in the example of the $v=3$ peak normal population model. The detection of such a higher peak, or retrieved populations incongruous with the Boltzmann distribution, in observed data would be consistent with a non-LTE signature. 

To facilitate the individual vibrational band detection, an `unpeeling' model is suggested, where individual band contributions are added to or subtracted from the total cross sections. 
Applying an unpeeling approach to the recent high-resolution detection of individual bands of OH in WASP-33b, we obtained a significant difference when band 1 was added to a single band 0 model illustrating the potential of detection of individual spectroscopic bands.

\section*{Acknowledgements}
This research is based [in part] on data collected at the Subaru Telescope, which is operated by the National Astronomical Observatory of Japan. We are honored and grateful for the opportunity of observing the Universe from Maunakea, which has the cultural, historical, and natural significance in Hawaii. Our data reductions benefited from (PyRAF and) PyFITS that are the products of the Space Telescope Science Institute, which is operated by AURA for NASA. We are also grateful to the developers of the {\sc Numpy}, {\sc Scipy}, {\sc Matplotlib}, {\sc Jupyter Notebook}, and {\sc Astropy} packages, which were used extensively in this work \citep{2020SciPy-NMeth, Hunter:2007, Kluyver:2016aa, astropy:2013, astropy:2018}.
{\it Funding:} SW was supported by the STFC UCL Centre for Doctoral Training in Data Intensive Science (grant number ST/P006736/1)
This project received funding from the European Research Council (ERC) under the European Union's Horizon 2020 research and innovation programme (grant agreements 758892, ExoAI and 883830, ExoMolHD) and the European Union's Horizon 2020 COMPET programme (grant agreement No 776403, ExoplANETS A). Furthermore, we acknowledge funding by the UK Space Agency and Science and Technology Funding Council (STFC) grants:  ST/R000476/1, ST/K502406/1, ST/P000282/1, ST/P002153/1, ST/S002634/1 and ST/T001836/1. The authors acknowledge the use of the UCL Legion High Performance Computing Facility (Legion@UCL) and associated support services in the completion of this work, along with the Cambridge Service for Data Driven Discovery (CSD3), part of which is operated by the University of Cambridge Research Computing on behalf of the STFC DiRAC HPC Facility (www.dirac.ac.uk). The DiRAC component of CSD3 was funded by BEIS capital funding via STFC capital grants ST/P002307/1 and ST/R002452/1 and STFC operations grant ST/R00689X/1. DiRAC is part of the National e-Infrastructure. 
This work was supported by JSPS KAKENHI grant Nos. JP22K14092, JP19K14783, JP21H00035, JP20H00170, 21H04998, and SATELLITE Research from the Astrobiology center (AB022006). JLB acknowledges funding from the European Research Council (ERC) under the European Union’s Horizon 2020 research and innovation program under grant agreement No 805445.

\section*{Data Availability}

 The underlying molecular linelist data from the ExoMol project \citep{20TeYuAl} are available from the Exomol website (\url{www.exomol.com}). Derived cross section data is available from the corresponding author upon request.

\newpage

\bibliography{main}{}
\bibliographystyle{aasjournal}



\end{document}